\documentclass[DIV=calc, paper=letter, fontsize=11pt]{scrartcl}	 

\usepackage[]{units}
\usepackage{lipsum} 
\usepackage[english]{babel} 
\usepackage[protrusion=true,expansion=true]{microtype} 
\usepackage{amsmath,amssymb,amsfonts,amsthm} 
\usepackage[svgnames, table]{xcolor} 
\usepackage[hang, small,labelfont=bf,up,textfont=it,up]{caption} 
\usepackage{booktabs} 
\usepackage{fix-cm}	 
\usepackage[]{units}

\usepackage{tabularx} 

\usepackage[colorlinks=true, linkcolor=blue, citecolor=blue, filecolor=blue, runcolor=blue, urlcolor = blue]{hyperref}

\usepackage{fullpage}
\usepackage{sectsty} 

\usepackage{fancyhdr} 
\usepackage{lastpage} 

\usepackage{nicefrac}
\usepackage{cite}

\usepackage{tikz}
\usetikzlibrary{calc,patterns,decorations.pathmorphing,decorations.markings}
\usepackage{pgfplots}

\usepackage{multicol}
\usepackage{float} 

\newcounter{tempcolnum}
\makeatletter
\newcommand{\multicolinterrupt}[1]{
\setcounter{tempcolnum}{\col@number}
\end{multicols}
#1%
\begin{multicols}{\value{tempcolnum}}
}
\makeatother

\usepackage{graphicx}
\DeclareGraphicsExtensions{.png}

\usepackage{lettrine} 
\newcommand{\initial}[1]{ 
\lettrine[lines=3,lhang=0.3,nindent=0em]{
\color{DarkGoldenrod}
{\textsf{#1}}}{}}

\usepackage{caption}

\usepackage{graphicx}

\usepackage{titling} 

\newcommand{\HorRule}{\color{DarkGoldenrod} \rule{\linewidth}{1pt}} 

\pretitle{\vspace{-80pt} \begin{flushleft} \HorRule \fontsize{20}{20} \color{DarkRed} \selectfont} 
\title{Hydrodynamic Coupling of Particle Inclusions Embedded in Curved Lipid Bilayer Membranes} 
\posttitle{\par\end{flushleft}} 
\preauthor{\vspace{-5pt} \begin{flushleft}\fontsize{14}{14} \large  \color{DarkRed}} 
\author{Jon Karl Sigurdsson and Paul J. Atzberger$^{*}$ } 
\postauthor{\fontsize{12}{12} \small \color{Black} 
Department of Mathematics, University of California Santa Barbara; e-mail: atzberg@math.ucsb.edu; website: http://atzberger.org/ \\
\vskip 1.5em
\lfoot{\footnotesize * Work supported by DOE ASCR CM4 DE-SC0009254 and NSF CAREER
Grant DMS-0956210.}
\end{flushleft} \vspace{-18pt} \HorRule} 

\DeclareMathOperator{\sech}{sech}

\newcommand{\LRratio}{\Pi_1}
\newcommand{\LRratioPlus}{\Pi_1^+}
\newcommand{\LRratioMinus}{\Pi_1^-}

\newcommand{\LRratioExp}{L/R}

\newcommand{\gammaMuRatio}{\Pi_2}
\newcommand{\gammaMuRatioPlus}{\Pi_2^+}
\newcommand{\gammaMuRatioMinus}{\Pi_2^-}
\newcommand{\gammaMuRatioExp}{\gamma R/\mu_f}

\newcommand{\mb}[1]{\mathbf{#1}}
\newcommand{\bs}[1]{\boldsymbol{#1}}

\newcommand{\subtxt}[1]{ {\mbox{\tiny #1}} }

\newcommand{\mDiv}{ {\mbox{div}} }
\newcommand{\mGrad}{ {\mbox{grad}} }
\newcommand{\mCurl}{ {\mbox{curl}} }

\definecolor{issuePJA_color}{rgb}{1.0,0.0,0.0}

\definecolor{commentPJA_color}{rgb}{1.0,0.0,0.8}

\begin{document}

\maketitle 

\thispagestyle{fancy} 


\initial{W}\textbf{e develop theory and computational methods to investigate particle inclusions embedded within curved lipid bilayer membranes.  We consider the case of spherical lipid vesicles where inclusion particles are coupled through (i) intramembrane hydrodynamics, (ii) traction stresses with the external and trapped solvent fluid, and (iii) intermonolayer slip between the two leaflets of the bilayer.  We investigate relative to flat membranes how the membrane curvature and topology augment hydrodynamic responses.  We show how both the translational and rotational mobility of protein inclusions are effected by the membrane curvature, ratio of intramembrane viscosity to solvent viscosity, and intermonolayer slip.  For general investigations of many-particle dynamics, we also discuss how our approaches can be used to treat the collective diffusion and hydrodynamic coupling within spherical bilayers.
}


\begin{multicols}{2}

\section{Introduction}
Cellular membranes are complex heterogeneous materials consisting of mixtures of lipids, proteins, and other small molecules~\cite{Alberts2007}.  The individual and collective dynamics of these species are fine-tuned to carry out complex cellular processes ranging from cell signalling to shape regulation of organelles~\cite{Alberts2007,Voeltz2007,Groves2007,Powers2002,Muller2012,NelsonStatMechMem2004}.  The effective two dimensional fluid-elastic nature of cell membranes results in interfacial phenomena and interesting geometric shapes effecting both molecular interactions and dynamics that can be very distinct from their bulk counter-parts.  To gain a deeper understanding of cellular processes requires insights into the fundamental mechanics of fluid-elastic bilayer membranes.

Early theoretical investigations of the hydrodynamics of flat lipid bilayer membranes include~\cite{Saffman1975,Saffman1976} and more recently the related works~\cite{Oppenheimer2009,Camley2013,Camley2012,LevineMobilityExtendedBodies2004,Naji2007c,Powers2002,Muller2012}.  In the now classic papers of Saffman and Delbr\"uck~\cite{Saffman1975,Saffman1976}, the bilayer is treated as a two dimensional fluid.  The two dimensional fluid is coupled to a bulk three dimensional fluid accounting for the solvent surrounding the membrane on both sides.  This description of the hydrodynamics is used to model a protein inclusion within a flat infinite membrane to derive the self-mobility 
$M_{SD} = (1/4\pi\mu_m)\left(\log({2L_{SD}}/a) - \gamma \right)$.  This asymptotic result assumes $a \ll L_{SD}$, where $a$ is the protein size, $\gamma \sim 0.577$ is the Euler-Mascheroni constant. The $L_{SD} = \mu_m/2\mu_f$ is the Saffman-Delbr\"uck length associated with how dissipation within the entrained bulk solvent fluid of viscosity $\mu_f$ regularizes the long-range two dimensional flow of viscosity $\mu_m$.  These results highlight the importance of dissipation in the bulk solvent fluid that if neglected would otherwise lead to the well-known Stokes paradox~\cite{Saffman1976,Lamb1895, HappelBrenner1983}.  This shows that particle motions even within a flat interface has a very different character than its counter-part in a bulk fluid. From Stokes theory the bulk self-mobility of a particle scales like $M ~\sim 1/6\pi\mu_f a$~\cite{Acheson1990,HappelBrenner1983}.  For curved membranes the topology and geometry can result in even more significant differences.  This includes providing a finite closed membrane surface and trapped solvent fluid in a bounded interior domain augmenting the hydrodynamics and coupling.

More recent works explore the mechanics of membranes both through coarse-grained molecular models~\cite{Deserno2009, Reynwar2007, Cooke2005, DesernoJanuary2015, Farago2003, Tieleman1997, Tieleman2013,Ayton2006} and through continuum mechanics approaches~\cite{Seifert1993, Seifert1997, DesernoJanuary2015, Capovilla2002, Camley2012, AtzbergerSigurdsson2012, AtzbergerBassereau2014, LevineMobilityExtendedBodies2004, LevineViscoelastic2002, LevineDinsmoreHydroEffectTopology2008, LevineHenleHydroCurvedMembranes2010, ArroyoRelaxationDynamics2009, Oppenheimer2009, Vlahovska2011, Chou2008, Klug2006,Powers2002, Muller2012, Ayton2006, OsterSteigmann2013, Lowengrub2007,Du2004}.  The particular works~\cite{LevineHenleHydroCurvedMembranes2010, LevineMobilityExtendedBodies2004, LevineDinsmoreHydroEffectTopology2008, Powers2002, NelsonStatMechMem2004, Noguchi2004} introduce a continuum mechanics description for the hydrodynamics of spherical vesicles and tubules.  The self-mobility of an embedded particle is computed as the curvature is varied using a truncation of the series representation of the hydrodynamic flow.  In~\cite{ArroyoRelaxationDynamics2009} an exterior calculus description of the continuum mechanics of a fluid-elastic membrane sheet is introduced and used to investigate lipid flow during processes such as budding with an asymptotic model for the ambient fluid.  The prior work in this area primarily has focused on single particle mobility and transport by hydrodynamics averaged over the two bilayer leaflets.

We introduce here further approaches to investigate the collective hydrodynamic coupling of multiple particle inclusions within leaflets of curved fluid lipid bilayer membranes.  We consider the case of spherical bilayer membranes where inclusion particles are coupled through (i) intramembrane hydrodynamics, (ii) traction stresses with the external and trapped solvent fluid, and (iii) intermonolayer slip between the two leaflets of the bilayer. We formulate tractable descriptions of the continuum mechanics of curved fluid bilayers drawing on results from the exterior calculus of differential geometry.  We formulate a tractable description for the collective hydrodynamic coupling of the inclusion particles on curved manifolds building on our prior work on immersed boundary approximations
~\cite{Atzberger2006,Atzberger2007a,AtzbergerSELM2011,AtzbergerSigurdsson2012}.  
We compute the translational and rotational mobilities of inclusion particles.  Relative to infinite flat membranes, we show that spherical vesicles exhibit significant differences arising from the curvature and finite domain size.  

\lfoot{} 

In Section~\ref{sec:ContMechVesicle} we introduce our continuum mechanics description of the bilayer hydrodynamics expressed in terms of the operators of exterior calculus of differential geometry.  We use exterior calculus to help take a less coordinate-centric approach in our derivations and to obtain more concise expressions that often have a more clear geometric interpretation.  We also show how the exterior calculus can be used to generalize many of the techniques used in fluid mechanics to the context of curved surfaces.  In Section~\ref{sec:LambSol}, we use Lamb's solution for the fluid flow exterior and interior to a spherical shell to obtain the traction stresses arising from the surrounding solvent fluid and the trapped solvent fluid.  In Section~\ref{sec:SphResponses}, we consider the hydrodynamic flow within the lipid bilayer membrane.  We use a spherical harmonics representation to derive analytic results for the solutions of the coupled hydrodynamic equations.  In section~\ref{sec:curvatureAndShear}, we discuss some roles played by curvature in hydrodynamic flows within surfaces.  

In Section~\ref{sec:particleBilayerCoupling}, we introduce immersed boundary approximations on manifolds to account for the coupling between the lipid flow and inclusion particles.  We discuss some particular properties of this type of approximation. We then derive mobility tensors for the translational and rotational motions of inclusion particles within curved membranes.  

In Section~\ref{sec:particleMobilities}, we investigate the self mobility and the coupled mobility of inclusion particles when varying (i) vesicle curvature, (ii) membrane viscosity vs solvent viscosity, and (iii) intermonolayer slip.  In Section~\ref{sec:manyParticleDynamics}, we consider approaches for the collective dynamics of many coupled inclusion particles within spherical vesicles.  We consider the collective mobility associated with an attracting cluster of particles and briefly discuss some of the interesting dynamics that can arise from the collective hydrodynamic coupling.  In Appendix~\ref{sec:coord_charts}, we discuss briefly how we have addressed some of the issues that arise for spherical surfaces in practical numerical calculations to obtain our results.  

In summary, the work presented here is meant as a starting point to understanding the basic features of the mobility of inclusion particles within curved bilayers.  Overall, we expect our approaches introduced here to provide ways to investigate the general collective dynamics of inclusion particles within spherical lipid bilayers relevant in many applications.

\section{Continuum Mechanics of the Vesicle}
\label{sec:ContMechVesicle}
\begin{figure}[H]
\centering
\includegraphics[width=8cm]{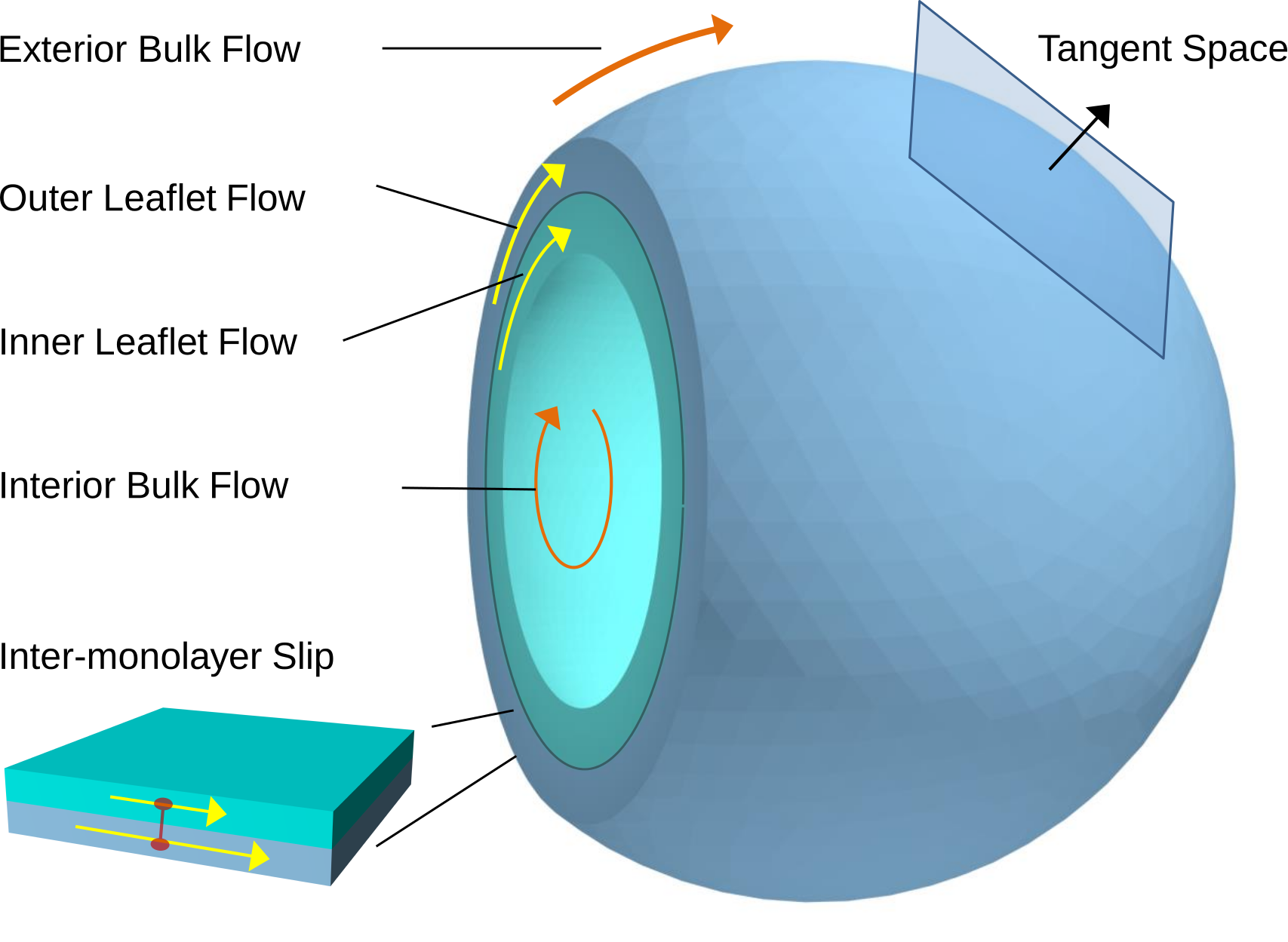}
\caption{Vesicle Hydrodynamics. We take into account the hydrodynamics within each leaflet of the bilayer, the intermonolayer slip between leaflets, and the traction stresses for both the solvent fluid trapped interior to the vesicle and the solvent fluid exterior to the vesicle.  We use a covariant formulation of the continuum mechanics.
\label{fig:hydroSchematic}}
\end{figure}

We formulate a continuum mechanics description of (i) the hydrodynamic flow of lipids within the two bilayer leaflets, (ii) intermonolayer slip, and (iii) coupling to the surrounding solvent fluid, see Figure~\ref{fig:hydroSchematic}.  We derive a set of conservation laws on manifolds using tensor calculus and results from differential geometry similar to Marsden~\cite{Marsden1994}.  We then use identities as in Arroyo and Disomone~\cite{ArroyoRelaxationDynamics2009} to express our equations in a convenient covariant form that is geometrically invariant.  To obtain analytic results for hydrodynamic flows on the curved surface, we use exterior calculus to generalize techniques often employed in fluid mechanics to 2-manifolds.  We then use these exterior calculus approaches to perform numerical calculations.  While our approaches provide rather general methods for working with hydrodynamics within manifolds, we focus in this paper on the sphere which is relevant to flow within the lipid bilayers of vesicles.

\subsection{Hydrodynamics of Bilayer Leaflets}
\label{sec:HydroBilayers}
We first consider the hydrodynamics within a single bilayer leaflet of the membrane.  We treat the membrane as a two-dimensional embedded continuum in the case that the surface velocity $\mb{V} = \mb{v} + v_n \mb{n}$ has zero velocity in the direction of the surface normal $v_n = 0$.  The conservation of momentum and mass of such a deforming two-dimensional continuum is given by~\cite{Marsden1994}
\begin{eqnarray}
\label{equ_cont_mech_gen}
\rho\left( \partial_t \mb{v} + \mb{v}\cdot \nabla \mb{v}\right)
& = &  \mDiv(\bs{\sigma}) + \mb{b} \\
\partial_t \rho + \rho \mDiv (\mb{v}) & = & 0.
\end{eqnarray}
The $\nabla$ denotes the covariant derivative which when expressed in terms of tensor components is $\left(\nabla \mb{v}\right)_{b}^a = v^a_{|b} =  \partial_{\mb{x}^b} v^a + \Gamma_{bc}^a v^c,$ where $\Gamma_{bc}^a$ denotes the Christoffel symbols~\cite{Pressley2001, Abraham1988}.   In the notation $\mDiv(\cdot)$ and $\mGrad(\cdot)$ the corresponding covariant operations for divergence $\mDiv(\mb{w}) = w_{|a}^a$ and gradient $\mGrad(\mb{w})^a_b = w_{|b}^a$.  The $\rho$ denotes the mass density per unit surface area, the $\mb{v}$ the velocity components tangent to the surface, $\mb{b}$ the body force per unit surface area, and $\bs{\sigma}$ the surface stress tensor.  We remark that while these equations look superficially similar to the Euclidean case owing to the convenient covariant derivative notation,
as we shall discuss the geometry introduces important differences and additional terms.

The constitutive law for an incompressible Newtonian fluid can be expressed in terms of the rate of deformation tensor of the surface
\begin{eqnarray}
\mb{D} = \nabla \mb{v} + \nabla \mb{v}^T,
\end{eqnarray}
which in terms of tensor components is $D^{a}_b = v^a_{|b} + v^b_{|a}$.
The Newtonian stress is given by 
\begin{eqnarray}
\bs{\sigma}^{\sharp} = \mu_m\mb{D}^{\sharp} +  \mu_m' \mDiv(\mb{v})\mathcal{I}^{\sharp}  - p \mathcal{I}^{\sharp}.
\end{eqnarray}
The $\mu_m$ and $\mu_m'$ are the first and second viscosities of the membrane.  The $\mathcal{I}$ is the (1,1)-identity tensor with $(\mathcal{I})_b^a = \delta_b^a$  where $\delta_b^a$ denotes the Kronecker delta-function.
This has $\left(\mathcal{I}^{\sharp}\right)^{ab} = g^{ab} = \left(\mb{g}^{\sharp}\right)^{ab}$, where $\mb{g}$ is the metric tensor for the surface ~\cite{Marsden1994}.  

We have adopted the notation for raising and lowering indices corresponding to the isomorphisms between the tangent and co-tangent spaces of the surface given by
\begin{eqnarray}
{\flat} & : &   v^j \partial_{\mb{x}^j} \rightarrow 
v_i d\mb{x}^i \\
{\sharp} & : &  v_i d\mb{x}^i \rightarrow v^j \partial_{\mb{x}^j}.
\end{eqnarray}
The $\partial_{\mb{x}^j}$ denotes the coordinate associated basis vectors of the tangent space and $d{\mb{x}^j}$ the one-form coordinate associated basis of the co-tangent space.  The isomorphisms can also be expressed directly in terms of the components as $v_i = g_{ij} v^j$ and $v^i = g^{ij} v_j$, where we denote the metric tensor as $g_{ij}$ and its inverse as $g^{ij}$ ~\cite{Abraham1988}.  This extends naturally to tensors.

Using these conventions, the steady-state Stokes equations corresponding to equation~\ref{equ_cont_mech_gen} can be expressed in tensor components as
\begin{eqnarray}
\mu_m D^{ab}_{|b} - g^{ab}p + {b}^a & = & 0 \\
v^a_{|a} & = & 0.
\end{eqnarray}
We can express this in a more geometrically transparent manner by considering further the divergence of the rate of deformation tensor 
\begin{eqnarray}
D^{ab}_{|b} & = & g^{ac}g^{bd} v_{c|d|b} + g^{ac}g^{bd} v_{d|c|b}.
\end{eqnarray}
We have that 
\begin{eqnarray}
g^{ac}g^{bd} v_{c|d|b} = ({\Delta}^R\mb{v})^a + Kv^a
\end{eqnarray}
where ${\Delta}^R \mb{v} := \mbox{rough-laplacian}( \mb{v}) = \mDiv\left(\mGrad\left(\mb{v}\right)\right)$ and 
$K$ is the Gaussian curvature of the surface.
We have that 
\begin{eqnarray}
g^{ac}g^{bd} v_{d|c|b} & = & \mGrad\left(\mDiv\left(\mb{v}\right)\right) + Kv^a \\
\nonumber
& = & Kv^a.
\end{eqnarray}
This follows since $\mDiv\left(\mb{v}\right) = 0$.

It is convenient to express the equations and differential operators in terms of the exterior calculus as follows.  Let $\mb{d}$ denote the usual exterior derivative for a $k$-form $\alpha$ as 
\begin{eqnarray}
\mb{d} \alpha = 
\frac{1}{k!}
\frac{\partial \alpha_{i_1 \ldots i_k}}{\partial x^j}  \mb{d}\mb{x}^j  \wedge \mb{d}\mb{x}^{i_1} \cdots \wedge \mb{d}\mb{x}^{i_k}.
\end{eqnarray}
The $\bs{\delta}$ denotes the co-differential operator given by 
$\bs{\delta} = \star \mb{d} \star$, where for a $k$-form $\alpha = \left(1/k!\right)\alpha_{i_1 \ldots i_k}\mb{d}\mb{x}^{i_1} \wedge \cdots \wedge \mb{d}\mb{x}^{i_{k}}$ the $\star$ denotes the Hodge star given by 
\begin{eqnarray}
\star \alpha = 
\frac{\sqrt{|g|}}{(n-k)!k!}
\alpha^{i_1 \ldots i_k}\epsilon_{i_1\ldots i_k j_1 \ldots j_{n-k}} \cdot \\
\hspace{1cm} \cdot \mb{d}\mb{x}^{j_1} \wedge \cdots \wedge \mb{d}\mb{x}^{j_{n-k}},
\end{eqnarray}
where 
$\alpha^{i_1 \ldots i_k} = g^{i_1 \ell_1}\cdots g^{i_k \ell_k} \alpha_{\ell_1 \ldots \ell_k}$,
$\sqrt{|g|}$ is the square-root of the determinant of the metric tensor, and 
$\epsilon_{i_1\ldots i_k j_1 \ldots j_{n-k}}$ denotes the 
Levi-Civita tensor~\cite{Marsden1994}.   

The generalization of the common differential operators of vector calculus to manifolds can be expressed in terms of exterior calculus as
\begin{eqnarray}
\mGrad(f)                 & = & \lbrack \mb{d}f\rbrack^{\sharp} \\
\mDiv(\mb{F}) & = & -(\star \mb{d}\star \mb{F}^\flat) = -\bs{\delta} \mb{F}^\flat  \\
\mCurl(\mb{F})        & = & \left\lbrack\star (\mb{d}\mb{F}^\flat) \right\rbrack^{\sharp}. 
\end{eqnarray}
The $f$ is a smooth scalar function and the $\mb{F}$ is a smooth vector field. 

There are different types of Laplacians that can be defined for manifolds
\begin{eqnarray}
\Delta^H (\mb{F}) 
& = & -\left\lbrack \left(\bs{\delta}\mb{d} + \mb{d}\bs{\delta} \right) \mb{F}^\flat\right\rbrack^{\sharp} \\
\Delta^S (\mb{F}) 
& = & -\left\lbrack \bs{\delta}\mb{d}  \mb{F}^\flat\right\rbrack^{\sharp} \\
\Delta^H f & = & \Delta^R f = -(\star \mb{d}\star )\mb{d} f = -\bs{\delta} \mb{d} f. 
\end{eqnarray}
The $\Delta^R = \mDiv(\mGrad(\cdot))$ denotes the rough-Laplacian given by the usual divergence of the gradient.  For vector fields, $\Delta^H (\mb{F})$ denotes the Hodge-de Rham Laplacian, which has similarities to taking the curl of the curl~\cite{Abraham1988}.  In fact, in the case that $\mDiv(\mb{F}) = -\bs{\delta} \mb{F}^{\flat} = 0$ we have $\Delta^H (\mb{F}) = \Delta^S (\mb{F}) = -\left[\bs{\delta} \mb{d} \mb{F}^{\flat}\right]^{\sharp}$.

Using these conventions, we have
\begin{eqnarray}
\mDiv(\mb{D}) & = & 
-\bs{\delta} \mb{d} \mb{v}^{\flat} + 2K\mb{v}^{\flat}
\end{eqnarray}
where we used that $\mDiv(\mb{v}) = -\bs{\delta} \mb{v}^{\flat} = 0$.
This allows for the steady-state Stokes problem on the surface to be expressed using exterior calculus in the convenient form
\begin{eqnarray}
\label{equ_Stokes_geometric}
\left\{
\begin{array}{llll}
\mu_m \left(-\bs{\delta} \mb{d} \mb{v}^{\flat} + 2 K \mb{v}^{\flat} \right) 
- \mb{d}p + \mb{b}^{\flat}
& = & 0 \\
-\bs{\delta} \mb{v}^{\flat} & = & 0.
\end{array}
\right.
\end{eqnarray}
As we shall discuss, this form provides a very convenient approach for analytic and numerical calculations.

\subsection{Coupling to External Solvent Fluid}
\label{sec:LambSol}

The solvent fluid surrounding the lipid bilayer membrane also exerts a traction stress on the inner and outer leaflets.   We account for this using the Stokes equations
\begin{eqnarray}
\label{equ_stokes_1}
\mu\Delta \mb{u} - \nabla p & = & 0, \hspace{0.3cm} \mbox{$\mb{x} \in \Omega$} \\
\label{equ_stokes_2}
\nabla \cdot \mb{u} & = & 0,  \hspace{0.3cm} \mbox{$\mb{x} \in \Omega$} \\
\label{equ_stokes_3}
\mb{u} & = & \mb{v},  \hspace{0.3cm} \mbox{$\mb{x} \in \partial \Omega$} \\
\label{equ_stokes_4}
\mb{u}_{\infty} & = & 0.
\end{eqnarray}
The $\Omega = \Omega^{\pm}$ denotes either the outside region $\Omega^{+}$ of fluid surrounding the vesicle or the domain 
$\Omega^{-}$ of fluid trapped inside the vesicle.

The solution to the Stokes equations and traction stress can be conveniently expressed in terms of harmonic functions.  This is most immediately seen for the pressure, which when taking the divergence of equation~\ref{equ_stokes_1}, yields 
\begin{eqnarray}
\Delta p & = & 0.
\end{eqnarray}
For the spherical geometry, the pressure can be expanded as
\begin{eqnarray}
p = \sum_{n = -\infty}^{\infty} p_n
\end{eqnarray}
where $p_n$ is the \textit{solid spherical harmonic} of order $n$
\begin{eqnarray}
\label{equ_q_ssph}
p_n(r,\theta,\phi) &=& 
r^n \sum_{m = -|n|}^{|n|} C_m^{n} Y_m^n(\theta,\phi) 
\end{eqnarray}
where 
\begin{eqnarray}
\label{equ_q_ssph_2}
Y_m^n(\theta,\phi) &=& e^{im\phi} P_n^{m}(\cos(\theta)).
\end{eqnarray}

For the solvent fluid velocity $\mb{u}^{-}$ in the domain $\Omega^{-}$ interior to the vesicle, 
Lamb showed that the solution can be expressed as
~\cite{HappelBrenner1983,Lamb1895}
\begin{eqnarray}
\mb{u}^{-} = \sum_{n = 1}^{\infty}  \mb{u}^{-}_n
\end{eqnarray}
where
\begin{eqnarray}
\mb{u}^{-}_n & = & \nabla \times \left(\mb{r}\chi_n\right) + \nabla \Phi_n \\ 
    & + & \frac{(n + 3)}{ 2\mu (n + 1)(2n + 3)} r^2\nabla p_n \\
   & - & \frac{n}{\mu(n + 1)(2n + 3)}\mb{r}p_n.
\end{eqnarray}
We shall refer to this as the \textit{Lamb's Solution}.
The surface flow $\mb{v}$ determines the solid spherical harmonic functions $\chi_n, \Phi_n,  p_n$ by the following relations 
\begin{eqnarray}
p_n    & = & \frac{\mu(2n + 3)}{n}\frac{1}{R} 
\left(\frac{r}{R}\right)^n \cdot \\
&\cdot & 
\nonumber
\left[
Y_n - (n - 1)X_n
\right] \\
\Phi_n & = & \frac{1}{2n} R \left(\frac{r}{R}\right)^n 
\left[
(n + 1)X_n - Y_n
\right] \\
\chi_n & = & \frac{1}{n(n + 1)}
\left(\frac{r}{R}\right)^n 
Z_n.
\end{eqnarray}
The $R$ is the radius of the spherical surface.
For the surface fluid velocity $\mb{V} = \mb{v} + v_n \mb{n}$ of the membrane, the $X_n, Y_n, Z_n$ are combined surface spherical harmonics of degree $n$ obtained by expanding the following scalar fields on the surface
\begin{eqnarray}
\label{equ_X_n}
\frac{\mb{r}}{r}\cdot \mb{V} & = & \sum_{n = -\infty}^{\infty} X_n \\
\label{equ_Y_n}
r \nabla \cdot \mb{V} & = & \sum_{n = -\infty}^{\infty} Y_n \\
\label{equ_Z_n}
\mb{r}\cdot \nabla \times \mb{V} & = & \sum_{n = -\infty}^{\infty} Z_n.
\end{eqnarray}
For the solvent fluid velocity $\mb{u}^{+}$ in the domain $\Omega^{+}$ exterior to the vesicle,
Lamb's solution is
~\cite{HappelBrenner1983,Lamb1895}
\begin{eqnarray}
\mb{u}^{+} = \sum_{n = 0}^{\infty}  \mb{u}^{+}_n
\end{eqnarray}
where
\begin{eqnarray}
\mb{u}^{+}_n & = & 
\nabla \times (\mb{r} \chi_{-(n+1)}) + \nabla \Phi_{-(n+1)}  \\
& - & \frac{(n-2)}{\mu 2n (2n - 1)} r^2 \nabla p_{-(n+1)}  \\
& + & \frac{(n + 1)}{\mu n (2n - 1)}\mb{r} p_{-(n+1)}.
\end{eqnarray}
The surface fluid velocity $\mb{V} = \mb{v} + v_n \mb{n}$ of the membrane, determines the harmonic functions 
$\chi_{-(n+1)}, \Phi_{-(n+1)}, p_{-(n+1)}$  giving
\begin{eqnarray}
p_{-(n+1)}    & = & \frac{\mu(2n - 1)}{n + 1}\frac{1}{R} 
\left(\frac{R}{r}\right)^{n +1} \cdot \\
\nonumber
&& \cdot \left[
(n + 2)X_n + Y_{n} 
\right] \\
\Phi_{-(n+1)} & = & \frac{1}{2(n + 1)} R \left(\frac{R}{r}\right)^{n + 1} \cdot \\
\nonumber
&& \cdot \left[
nX_n - Y_n
\right] \\
\chi_{-(n+1)} & = & \frac{1}{n(n + 1)}
\left(\frac{R}{r}\right)^{n+1}
Z_{n}. 
\end{eqnarray}
In the special case of $v_n = 0$ and $\mDiv(\mb{v}) = 0$ we have that $X_n = Y_n = 0$ and that only the $Z_n$ term is non-trivial.  The Lamb's solution simplifies to 
\begin{eqnarray}
\label{equ_lambs_sol}
\mb{u}^{+}_n & = & \nabla \times \left(\mb{r}\chi_{-(n+1)}\right) \\
\mb{u}^{-}_n & = & \nabla \times \left(\mb{r}\chi_n\right).
\end{eqnarray}
In this case, the traction stress of the external fluid on the lipid bilayer membrane is
\begin{eqnarray}
\label{equ_tract_stress_tplus}
\mb{t}^{+} & = & \bs{\sigma}^{+}\cdot \mb{n}^{+} \\
\nonumber
& = & \left[\mu_{+} 
\left(
\nabla \mb{u}^{+} + \nabla \mb{u}^{+T} 
\right)
- p^{+}\mathcal{I} \right]\cdot \mb{n}^{+} \\
\nonumber
& = &
\mu_{+}
\frac{\partial \mb{u}^{+}}{\partial r} 
 +
\mu_{+}
\nabla
\left(\mb{u}^{+}\cdot \mb{n}^{+}\right) \\
\label{equ_tract_stress_tminus}
\mb{t}^{-} & = & \bs{\sigma}^{-}\cdot \mb{n}^{-} \\
\nonumber
& = & \left[\mu_{-} 
\left(
\nabla \mb{u}^{-} + \nabla \mb{u}^{-T} 
\right)
- p^{-}\mathcal{I} \right]\cdot \mb{n}^{-} \\
\nonumber
& = & 
-\mu_{-}
\frac{\partial \mb{u}^{-}}{\partial r} 
 +
\mu_{-}
\nabla
\left(\mb{u}^{-}\cdot \mb{n}^{-}\right).
\end{eqnarray}
The $\mb{n}^{\pm}$ denotes the unit normal on the surface $\partial \Omega^{\pm}$ in the direction pointing into the domain.  In these expressions, we emphasize that $\mb{n}^{\pm}$ is to be held fixed during differentiation.  
From equation~\ref{equ_tract_stress_tplus} and~\ref{equ_tract_stress_tminus} and the properties of solid spherical harmonics, we have that the traction stress on the membrane leaflets can be expressed as
\begin{eqnarray}
\label{equ_tract_stress_u_pm}
\mb{t}^{+} & = & \sum_{n = 0}^{\infty} -\frac{(n + 2)}{R^{+}} \mb{u}_n^{+} \\
\nonumber
\mb{t}^{-} & = & \sum_{n = 1}^{\infty} -\frac{(n - 1)}{R^{-}} \mb{u}_n^{-}.
\end{eqnarray}

\subsection{Intermonolayer Slip}
We account for the two bilayer leaflets of the membrane by considering two surface velocity fields $\mb{v}_+$ and $\mb{v}_-$.  We model the intermonolayer slip between these two leaflets by the traction term proportional to the difference in lipid velocity
\begin{eqnarray}
\mb{s}^{\pm} &=& \pm \gamma\left(\mb{v}_-  - \mb{v}_+  \right).
\end{eqnarray}

\subsection{Full Membrane Hydrodynamics}
\label{sec:fullMemHydro}
Using an approach similar to Section~\ref{sec:HydroBilayers}, we obtain for a two-leaflet membrane the following hydrodynamic equations.
\begin{align}
\label{equ_full_hydro_first}
\\
\nonumber
\left\{  
\begin{array}{ll}
\mu_m 
\left[
-\bs{\delta} \mb{d} \mb{v}_{+}^{\flat}
+ 
2K_{+} \mb{v}_{+}^{\flat}
\right]
+ 
\mb{t}_{+}^{\flat}
-\gamma 
\left(
\mb{v}_{+}^{\flat} - \mb{v}_{-}^{\flat}
\right) \\
\hspace{0.5cm}
= 
\mb{d}p_{+} - \mb{b}_{+}^{\flat}
= 
-\mb{c}_{+}^{\flat},
\hspace{0.25cm}
\mb{x} \in \mathcal{M}_{+} \\
\bs{\delta} \mb{v}_{+}^{\flat} = 0, 
\hspace{2.64cm}
\mb{x} \in \mathcal{M}_{+}, \\
\nonumber
\\
\nonumber
\mu_m 
\left[
-\bs{\delta}\mb{d} \mb{v}_{-}^{\flat}
+ 
2K_{-} \mb{v}_{-}^{\flat}
\right]
+ 
\mb{t}_{-}^{\flat}
-\gamma 
\left(
\mb{v}_{-}^{\flat} - \mb{v}_{+}^{\flat}
\right) \\
\hspace{0.5cm}
= 
\mb{d}p_{-} - \mb{b}_{-}^{\flat}
= 
-\mb{c}_{-}^{\flat},
\hspace{0.25cm}
\mb{x} \in \mathcal{M}_{-} \\
\bs{\delta} \mb{v}_{-}^{\flat} = 0, 
\hspace{2.64cm}
\mb{x} \in \mathcal{M}_{-}.
\end{array}
\right. 
\label{equ_full_hydro_last}
\end{align}
The $\mathcal{M}_{\pm}$ denotes the two surfaces representing the inner and outer bilayer leaflets.  These equations take into account the internal membrane hydrodynamics of each leaflet of viscosity $\mu_m$, the intermonolayer slip $\gamma$, and the traction stresses with the bulk solvent fluids of viscosity $\mu^{\pm}$ trapped within and external to the vesicle.  To obtain the coupling in the collective dynamics of inclusions embedded in such bilayer membranes, we must solve these equations for the hydrodynamic flow. 

\subsection{Membrane Hydrodynamics and Modal Responses}
\label{sec:SphResponses}
We use exterior calculus methods to derive solutions to the membrane hydrodynamic equation~\ref{equ_full_hydro_first}.  For analytic and numerical calculations of flow within surfaces, the exterior calculus provides a number of advantages over more coordinate-centric approaches such as tensor calculus~\cite{Abraham1988}.  As already seen in our expressions of the hydrodynamic equations, there are fewer explicit references to the metric tensor with instead more geometrically intrinsic operations appearing such as the exterior derivative and Hodge star~\cite{Abraham1988}.  In analytic calculations, we take advantage of this to develop succinct methods for curved manifolds that generalize many of the vector calculus based techniques often employed in fluid mechanics.  From the exterior calculus formulation of the Stokes equations we can readily show that the incompressible surface flow can be expressed in terms of a scalar velocity potential $\Phi$ as
\begin{eqnarray}
\label{equ_gen_curl_2}
\mb{v}^{\flat} = -\star \mb{d} \Phi.
\end{eqnarray}
This provides a generalization for the surface geometry of the usual velocity potential used in fluid mechanics.  The equation~\ref{equ_gen_curl_2} generalizes to surfaces the operation in Euclidean space of taking the curl to obtain an incompressible flow~\cite{HappelBrenner1983,Acheson1990}.  The exterior calculus allows us to readily verify that the generated velocity field on the surface is incompressible 
\begin{eqnarray}
-\bs{\delta} \mb{v}^{\flat} = (\star \mb{d} \star)(\star \mb{d} \Phi) = -\star \mb{d}^2 \Phi = 0.
\end{eqnarray}
This follows since $\star\star = -1$ on a surface (2-manifold) and $\mb{d}^2 = 0$ holds~\cite{Abraham1988}.  

\subsubsection{Modal Response for Intramembrane Hydrodynamics}
\label{equ_modal_response_intramembrane}
To obtain equations for $\Phi$, we use the exterior calculus to determine the eigenfunctions of the operator in the Stokes equations. This can then be used to rigorously derive expressions for the modal responses of the hydrodynamics when acted upon by an applied force in a manner similar to~\cite{LevineHenleHydroCurvedMembranes2010}.  For this purpose, we consider the eigenproblem 
\begin{eqnarray}
\label{equ_eigenproblem}
\mu\left[
-\bs{\delta}\mb{d} \mb{v}_s^{\flat}
+ 2K\mb{v}_s^{\flat}
\right]
& = & \lambda_s \mb{v}_s^{\flat}.
\end{eqnarray}
Let $\Phi_s$ be a function such that 
$\mb{v}_s^{\flat} = -\star \mb{d} {\Phi}_s$.  The operator 
$-\star \mb{d}$ commutes with $-\bs{\delta}\mb{d}$ since 
$-\bs{\delta}\mb{d} \mb{v}_s^{\flat} = -\bs{\delta}\mb{d}(-\star \mb{d}) {\Phi}_s 
= \star \mb{d} \star \mb{d} \star \mb{d} \Phi_s
= -\star \mb{d}\left(-\bs{\delta}\mb{d} \right){\Phi}_s$.  The eigenproblem becomes 
\begin{align}
(-\star \mb{d})
&\mu\left[
-\bs{\delta}\mb{d} \Phi_s
+ 2K\Phi_s
\right] \\
\nonumber
&= 
(-\star \mb{d}) (\lambda_s \Phi_s).
\end{align}
This can be satisfied if $\Phi_s$ is a solution of
\begin{align}
\mu\left[
-\bs{\delta}\mb{d} \Phi_s
+ 2K\Phi_s
\right] 
= 
\lambda_s \Phi_s.
\end{align}
This can also be expressed as 
\begin{eqnarray}
\label{equ_Laplace_Beltrami}
-\bs{\delta}\mb{d} \Phi_s
& = &
\gamma_s \Phi_s,
\end{eqnarray}
where 
$\gamma_s = 
\left(
{\lambda_s}/{\mu}
- 2K
\right)
$.
For scalar fields, the operator $-\bs{\delta}\mb{d}$ is the Laplace-Beltrami operator of the surface.  In the special case of the sphere, the solutions are surface spherical harmonics of the form
\begin{eqnarray}
\Phi_s = Y_m^\ell(\theta,\phi) = e^{im\phi} P_{\ell}^{m}(\cos(\theta))
\end{eqnarray}
where $s = (\ell,m)$ subject to the restriction $|m| \leq \ell$.  The eigenvalues are $\gamma_s = -\ell (\ell + 1)/R^2$
and $\lambda_s =\mu\left(-\ell (\ell + 1)/R^2 + 2K   \right)$.  

We can express the solution of the Stokes equations~\ref{equ_Stokes_geometric} by expanding the velocity field as 
\begin{eqnarray}
\mb{v}^{\flat} = 
\sum_s a_s \mb{v}_s^{\flat} 
= -\star \mb{d}
\sum_s a_s \Phi_s.
\end{eqnarray}
We can also represent the solution with $\Phi = \sum_s a_s \Phi_s$.  In a similar manner, the applied surface force can be expanded with coefficients $c_s$ as $\mb{c}^{\flat} = \mb{b}^{\flat} - \mb{d}p = -\star \mb{d}
\sum_s c_s \Phi_s$.  The problem now becomes to find the coefficients $a_s$ for the flow when given an applied force with coefficients $c_s$.

As a demonstration of the utility of this exterior calculus approach, consider the 
Stokes equations~\ref{equ_Stokes_geometric} for the surface flow on a sphere associated with a single leaflet, without yet the intermonolayer slip or traction stress.  We treat the Stokes problem in equation~\ref{equ_Stokes_geometric} by taking $-\star \mb{d}$ of both sides to eliminate the pressure term.  This yields 
\begin{eqnarray}
-\star \mb{d}\mu\left[
-\bs{\delta}\mb{d} \mb{v}^{\flat}
+ 2K\mb{v}^{\flat}
\right]
-\star \mb{d}\mb{d}p
=  
-\star \mb{d} \mb{b}^{\flat}.
\end{eqnarray}
Using the expansion for $\mb{v}^{\flat}$ in terms of $\mb{v}_s^{\flat} = -\star \mb{d} \Phi_s$ and that $\Phi_s$ was chosen to solve the eigenproblem in equation~\ref{equ_eigenproblem}, we have 
\begin{align}
-\star \mb{d}&\mu\left[
-\bs{\delta}\mb{d} \mb{v}^{\flat}
+ 2K\mb{v}^{\flat}
\right]
- 0 \\
\nonumber
& =   
-\star \mb{d}  \sum_s a_s \lambda_s ( -\star\mb{d} \Phi_s) \\
\nonumber
&=
-\sum_s a_s \lambda_s (-\bs{\delta}\mb{d} \Phi_s) 
= -\sum_s a_s \lambda_s \gamma_s \Phi_s  
\\ 
\nonumber
&=
-\sum_s c_s \star \mb{d}\star \mb{d} \Phi_s 
= 
-\sum_s c_s \bs{\delta}\mb{d} \Phi_s \\
\nonumber
&=
\sum_s c_s \gamma_s \Phi_s.
\end{align}
We remark that we use $\mb{c}^{\flat}$ as opposed to $\mb{b}^{\flat}$ throughout our calculations to emphasize that only the solenoidal component of the applied force effects the flow.  This is further exhibited in the identity $-\star \mb{d} \mb{c}^{\flat} = -\star \mb{d} \mb{b}^{\flat}$.
For mode $s$, we have $\lambda_s a_s = c_s$ and $K = 1/R^2$ which gives
\begin{eqnarray}
a_s = \left[ \left( \frac{\mu (2 - \ell (\ell + 1))}{R^2} \right) \right]^{-1} c_s.
\end{eqnarray}
This applies for $\ell \geq 2$.  For the Stokes flow on the membrane surface this gives the modal response to an applied force.  

We have assumed for this solution that the applied force has net-zero torque.  The mode $\ell = 1$ does not introduce an internal shear stress within the membrane since this mode corresponds to a rigid-body motion of the spherical shell.  Since we have not yet included the intermonolayer slip or the external fluid traction stress there would be no stresses to balance a force having non-zero net torque.      

We remark that the membrane velocity field is obtained from these calculations by
\begin{eqnarray}
\label{equ_vel_rep}
\mb{v} & = & \left(\mb{v}^{\flat}\right)^{\sharp} = \sum_s a_s \left(-\star \mb{d} \Phi_s\right)^{\sharp} \\
\nonumber
& = & \sum_s a_s
\left[ \frac{\epsilon_{i\ell}}{\sqrt{|g|}} \frac{\partial \Phi}{\partial x^{\ell}} \right]\partial_{\mb{x}^{i}}.
\end{eqnarray}
The $|g| = \mbox{det}(\mb{g})$ is the determinant of the metric tensor and $\epsilon_{i\ell}$ is the Levi-Civita tensor (slight abuse of notation).  The $x^\ell$ denotes the coordinates.  

We remark if the standard spherical coordinates are used then $x^1 = \theta$ and $x^2 = \phi$ with the polar angle $\theta$ and azimuthal angle $\phi$.  However, this has singularities at the north and south poles of the sphere.  To use robustly the velocity representation~\ref{equ_vel_rep} for numerical calculations at each location on the sphere, we need to use different coordinate charts depending on location.  The velocity field can be computed robustly by using either the standard spherical coordinates or the spherical coordinates with the poles at the east and west poles, for details see our discussion in Appendix~\ref{sec:coord_charts}.

\subsubsection{Modal Response when Coupled to External Solvent Fluid and with Intermonolayer Slip}
Using this approach, we can also incorporate for a two leaflet lipid bilayer membrane the additional contributions of the traction stress with the external solvent fluid and the intermonolayer slip.  We consider the case when the outer bilayer leaflet of the membrane vesicle has radius $R_{+}$ and the inner bilayer leaflet has radius $R_{-}$.  This requires us to derive the modal response to the induced bulk external solvent flow.  The solvent flow satisfies the Stokes equations~\ref{equ_stokes_1}--\ref{equ_stokes_4} with no-slip with respect to the flow within the membrane surface.  These Stokes equations must be solved twice, once in the domain $\Omega^{+}$ exterior to $\mathcal{M}^{+}$ and once in the domain $\Omega^{-}$ interior to $\mathcal{M}^{-}$.  We obtain a representation for these solutions using Lamb's solution~\cite{HappelBrenner1983}, see Section~\ref{sec:LambSol}.  

We represent the fluid velocity $\mb{v}_{\pm}$ within each leaflet of the membrane using the velocity potential $\Phi^{\pm}$.  As in Section~\ref{equ_modal_response_intramembrane}, we expand the velocity potential in spherical harmonics $\Phi_s$ as $\Phi^{\pm} = \sum_s a_s^{\pm} \Phi_s$.  This allows us to express the membrane velocity as
\begin{eqnarray}
\mb{v}_{\pm}^{\flat} = 
-\star \mb{d}
\sum_s a_s^{\pm} \Phi_s.
\end{eqnarray}
From this representation and equation~\ref{equ_tract_stress_u_pm}, we can express the traction stress from the external solvent fluid on the membrane leaflet as
\begin{eqnarray}
\mb{t}_{+}^{\flat} & = & \sum_{\ell = 1}^{\infty} -\frac{\mu_{+}(\ell + 1)}{R_{+}} 
\left(-\mb{d}\star \tilde{\Phi}_\ell^{-}\right)\\
\nonumber
\mb{t}_{-}^{\flat} & = & \sum_{\ell = 1}^{\infty} -\frac{\mu_{-}(\ell - 1)}{R_{-}} 
\left(-\mb{d}\star \tilde{\Phi}_\ell^{+}\right).
\end{eqnarray}
The $\tilde{\Phi}_\ell^{\pm}$ denotes the linear combination of modes of degree $\ell$.  In particular, $\tilde{\Phi}_\ell^{\pm} = \sum_{s', \ell' = \ell} a_{s'}^{\pm} \Phi_{s'}$ where $s' = (m',\ell')$.

By applying $-\star \mb{d}$ we have
\begin{eqnarray}
\label{equ_star_d_t}
-\mb{d}\star \mb{t}_{+}^{\flat} & = & -\frac{\mu_{+}}{R_{+}} 
\sum_{s} (\ell + 1)
a_s^{+} \gamma_s^{+} \Phi_s^{+} \\
\nonumber
-\mb{d}\star \mb{t}_{-}^{\flat} & = & -\frac{\mu_{-}}{R_{-}} 
\sum_{s} (\ell - 1)
a_s^{-} \gamma_s^{-} \Phi_s^{-}.
\end{eqnarray}
We use that $-\star\mb{d}\left(-\star\mb{d}\right) = -\bs{\delta}\mb{d}$ is the Laplace-Beltrami operator and that $-\bs{\delta}\mb{d} \Phi_s^{\pm} = \gamma_s^{\pm} \Phi_s^{\pm}$, see equation~\ref{equ_eigenproblem} and equation~\ref{equ_Laplace_Beltrami}.

Now we apply the operator $-\star \mb{d}$ to equation~\ref{equ_full_hydro_first}. By using equation~\ref{equ_star_d_t} and equation~\ref{equ_eigenproblem}, we obtain for the coefficients $a_s^{\pm}$ of the velocity fields of the leaflets
\begin{eqnarray}
\mu_m \lambda_s^{+} \gamma_s^{+} a_{s}^{+} 
& - &\frac{\mu_{+}}{R_{+}}(\ell +1) \gamma_s^{+} a_{s}^{+} \\
\nonumber
& - & \gamma \left(\gamma_s^{+} a_{s}^{+} - \gamma_s^{+} a_{s}^{-} \right)  
= -\gamma_s^{+} c_{s}^{+} \\
\mu_m \lambda_s^{-} \gamma_s^{-} a_{s}^{-} 
& - &\frac{\mu_{-}}{R_{-}}(\ell-1) \gamma_s^{-} a_{s}^{-} \\
\nonumber
& - & \gamma \left(\gamma_s^{-} a_{s}^{-} - \gamma_s^{-} a_{s}^{+} 
\right) = -\gamma_s^{-} c_{s}^{-}. 
\end{eqnarray}
The solution coefficients for $\mb{v}_{+}^{\flat}$ and $\mb{v}_{-}^{\flat}$ for the full two-leaflet membrane hydrodynamics in equation~\ref{equ_full_hydro_first} can be expressed as 
\begin{eqnarray}
\label{equ_Stokes_full_1}
\left[
\begin{array}{l}
a^{+}_s \\
a^{-}_s \\
\end{array}
\right]
& = & \mathcal{A}_s^{-1} 
\left[
\begin{array}{l}
-c_s^{+} \\
-c_s^{-} \\
\end{array}
\right]
\end{eqnarray}
where 
\begin{eqnarray}
\label{equ_Stokes_SPH_sol2_defAs}
\mathcal{A}_s & = & 
\left[
\begin{array}{ll}
A_1^{\ell} 
 - \gamma & \gamma \\
\gamma &
A_2^{\ell} - \gamma \\
\end{array}
\right]
\end{eqnarray}
with 
\begin{eqnarray}
\label{equ_Stokes_SPH_sol2_defA_ells} 
\\
\nonumber
A_1^{\ell} & = & 
\frac{\mu_m}{R_{+}^2}
\left(
2 - \ell(\ell + 1) - \frac{R_{+}}{L^{+}}(\ell + 1) 
\right)  \\
\nonumber
A_2^{\ell} & = & 
\frac{\mu_m}{R_{-}^2}
\left(
2 - \ell(\ell + 1) - \frac{R_{-}}{L^{-}}(\ell - 1) 
\right).
\end{eqnarray}
Associated with the inner and outer external fluids, we define the length-scales
$L^{-} = \mu_m/\mu_{-}$ and $L^{+} = \mu_m/\mu_{+}$.  The Saffman-Delbr\"uck length-scale~\cite{Saffman1975,Saffman1976} associated with each leaflet is
$L_{SD}^{-} = \frac{1}{2}L^{-}$ and $L_{SD}^{+} = \frac{1}{2}L^{+}$ and on average $L_{SD} = \frac{1}{2}\left(L_{SD}^{-} + L_{SD}^{+} \right)$. 

In summary, the equations~\ref{equ_Stokes_full_1}--~\ref{equ_Stokes_SPH_sol2_defA_ells} provide the modal responses for the hydrodynamic flow satisfying the two leaflet Stokes problem in equation~\ref{equ_full_hydro_first}.  The model captures the hydrodynamic flow of lipids within the two curved bilayer leaflets that are coupled to one another by intermonolayer slip and that are coupled to the flow of the external solvent fluid.  The key parameters are given in Table~\ref{table:descrParams}.

We remark that the membrane fluid velocity fields are obtained from these calculations by
\begin{eqnarray}
\label{equ_velFieldRep}
\\
\nonumber
\mb{v}_{-} & = & \left(\mb{v}_{-}^{\flat}\right)^{\sharp} = \sum_s a_s^{-} \left(-\star \mb{d} \Phi_s\right)^{\sharp} \\
& = & 
\nonumber
\sum_s a_s^{-}
\left[ \frac{\epsilon_{i\ell}}{\sqrt{|g_{-}|}} \frac{\partial \Phi_s}{\partial x^{\ell}} \right]\partial_{\mb{x}^{i}} \\
\nonumber
\mb{v}_{+} & = & \left(\mb{v}_{+}^{\flat}\right)^{\sharp} = 
\sum_s a_s^{+} \left(-\star \mb{d} \Phi_s \right)^{\sharp} \\
& = & 
\nonumber
\sum_s a_s^{+}
\left[ \frac{\epsilon_{i\ell}}{\sqrt{|g_{+}|}} \frac{\partial \Phi_s}{\partial x^{\ell}} \right]\partial_{\mb{x}^{i}}. 
\end{eqnarray}
The $|g_{\pm}|$ denotes the determinant of each of the metric tensors $\mb{g}_{\pm}$ associated with the leaflet surfaces $\mathcal{M}^{\pm}$ and the $\epsilon_{i\ell}$ denotes the Levi-Civita tensor. We remark that given coordinate singularities on the sphere to use robustly this velocity representation for numerical calculations, we need to use different coordinate charts.  For details see our discussion in Appendix~\ref{sec:coord_charts}.

\begin{table}[H]
\centering
\begin{tabular}{|l|l|}
\hline
\rowcolor{LightGrey}
Notation & Description \\
\hline
$\mu_m$ & \small Intramembrane viscosity. \\ 
$\gamma$ & \small Intermonolayer slip. \\ 
$\mu_{\pm}$ & \small External bulk solvent viscosity. \\ 
$R_\subtxt{+}$ & \small Radius of the outer leaflet. \\
$R_\subtxt{-}$ & \small Radius of the inner leaflet. \\ 
$R$ & \small Average radius of the vesicle. \\ 
\hline 
\end{tabular}
\caption{Description of notation and parameters.
\label{table:descrParams}}
\end{table}

\subsection{Characteristic Physical Scales}
\label{sec:charScales}
To characterize the hydrodynamic responses, we discuss a few useful non-dimensional groups.  We first consider how the bulk solvent fluid regularizes the two dimensional membrane hydrodynamics.  This can be characterized by considering a disk-shaped patch of a flat membrane of radius $r$.  An interesting length-scale is the radius $r$ where the bulk three dimensional traction stress acting on the patch of area $A = \pi r$ is comparable in magnitude to the intramembrane stresses acting on the perimeter of the patch of length $\tilde{\ell} = 2\pi r$.  This occurs for the inner and outer leaflets on length-scales scaling respectively like $L^{-} = \mu_m/\mu_{-}$ and $L^{+} = \mu_m/\mu_{+}$.  The Saffman-Delbr\"uck length-scale~\cite{Saffman1975,Saffman1976} associated with each leaflet is
$L_{SD}^{-} = \frac{1}{2}L^{-}$ and $L_{SD}^{+} = \frac{1}{2}L^{+}$ with average $L_{SD} = \frac{1}{2}\left(L_{SD}^{-} + L_{SD}^{+} \right)$.  For a vesicle, it is natural to consider these length-scales relative to the radius of the vesicle $R$.  We introduce the non-dimensional groups $\LRratioPlus = L^{+}/R_{+}$ and  $\LRratioMinus = L^{-}/R_{-}$.

For the intermonolayer slip, we consider for the flow the response of the leading order modes with $\ell = 1$.  These correspond to the rigid rotations of the spherical shell.  For a velocity difference between the layers, the drag is given by $\gamma$.  For the leading order modes with $\ell = 1$, the traction stresses arising from the entrained surrounding bulk solvent fluid give an effective drag ${\mu_{+}}/{R_{+}}$, see equation~\ref{equ_Stokes_SPH_sol2_defAs} and ~\ref{equ_Stokes_SPH_sol2_defA_ells}.  To characterize for a lealfet the strength of the intermonolayer slip relative to the traction stress exerted by the surrounding solvent fluid, we introduce the
non-dimensional groups $\gammaMuRatioPlus = \gamma R_{+}/\mu_{+}$ and $\gammaMuRatioMinus = \gamma R_{-}/\mu_{-}$.  For convenience, we also introduce the notation $\gamma^{\pm}_0 = \mu_{\pm}/R_{\pm}$, so that we can express $\gammaMuRatio^{\pm} = \gamma/\gamma^{\pm}_0$.

We remark that $\gammaMuRatio^{\pm}$ can be expressed in the more familiar terms of a ratio of rotational drag coefficients.  We have $\gammaMuRatioPlus = \left[8 \pi (\gamma R_{+}) R_{+}^3\right] / \left[8 \pi \mu_{+} R_{+}^3\right]$.  For a rigid spherical particle subject to torque $\tau$ in a fluid with viscosity $\bar{\mu}$, the angular velocity $\omega$ is given by $\omega = \left[8\pi\bar{\mu} R_{+}^3\right]^{-1} \tau$,~\cite{HappelBrenner1983}.  This shows that the intermonolayer slip contributes similarly to leading order as a bulk solvent fluid of viscosity $\gamma R_{+}$.  We can express similarly $\gammaMuRatioMinus$.  

These four non-dimensional groups $\LRratioPlus$, $\LRratioMinus$, $\gammaMuRatioPlus$, $\gammaMuRatioMinus$ serve to characterize the relative contributions of the vesicle geometry, shear viscosity within the bilayer leaflets, the shear viscosity of the bulk solvent fluid, and the intermonolayer slip.  To simplify our notation, we drop the $\pm$ when the same values are used for each leaflet and denote $\Pi_1 = \LRratioPlus = \LRratioMinus$ and $\Pi_2 = \gammaMuRatioPlus = \gammaMuRatioMinus$.  

We can non-dimensionalize equations~\ref{equ_Stokes_full_1_nonDim}-- \ref{equ_Stokes_SPH_sol2_defA_ells_nonDim} by introducing a characteristic velocity $v_0^{\pm}$ and force density $f_0^{\pm}$.  We find it convenient to consider the rigid rotation at angular velocity $\omega_0$ of the spherical shell in the solvent fluid.  This motivates the choice of characteristic force density $f_0^{\pm} = \mu_f \omega_0$ and velocity $v_0^{\pm} = R \omega_0$.  For simplicity, we consider only the case with $\mu^{\pm} = \mu_f$ and $R_{\pm} = R$.   The non-dimensional velocity is $\tilde{\mb{v}}_{\pm}^{\flat} = {\mb{v}}_{\pm}^{\flat}/v_0$ and force density $\tilde{\mb{c}}_{\pm}^{\flat} = {\mb{c}}_{\pm}^{\flat}/f_0$
with coefficients $\tilde{a}$ and $\tilde{c}$.
With this choice, we can express the non-dimensional problem for the full two-leaflet membrane hydrodynamics in equation~\ref{equ_full_hydro_first} as 
\begin{eqnarray}
\label{equ_Stokes_full_1_nonDim}
\left[
\begin{array}{l}
\tilde{a}^{+}_s \\
\tilde{a}^{-}_s \\
\end{array}
\right]
& = & \tilde{\mathcal{A}}_s^{-1} 
\left[
\begin{array}{l}
-\tilde{c}_s^{+} \\
-\tilde{c}_s^{-} \\
\end{array}
\right]
\end{eqnarray}
where 
\begin{eqnarray}
\label{equ_Stokes_SPH_sol2_defAs_nonDim}
\tilde{\mathcal{A}}_s & = & 
\Pi_2
\left[
\begin{array}{ll}
\tilde{A}_1^{\ell} 
 - 1 & 1 \\
1 &
\tilde{A}_2^{\ell} - 1 \\
\end{array}
\right]
\end{eqnarray}
with 
\begin{eqnarray}
\label{equ_Stokes_SPH_sol2_defA_ells_nonDim} 
\\
\nonumber
\tilde{A}_1^{\ell} & = & 
\Pi_2^{-1}\Pi_1
\left(
2 - \ell(\ell + 1) - \Pi_1^{-1}(\ell + 1) 
\right)  \\
\nonumber
\tilde{A}_2^{\ell} & = & 
\Pi_2^{-1}\Pi_1
\left(
2 - \ell(\ell + 1) - \Pi_1^{-1}(\ell - 1) 
\right).
\end{eqnarray}
As we shall discuss, this analysis will be useful when considering the relative contributions of the solvent traction stress, intramembrane viscosity, and the intermonolayer slip.  Other non-dimensional scalings can also be considered using a similar approach.

\subsection{Curvature and Shear}
\label{sec:curvatureAndShear}
\begin{figure}[H]
\centering
\includegraphics[width=8cm]{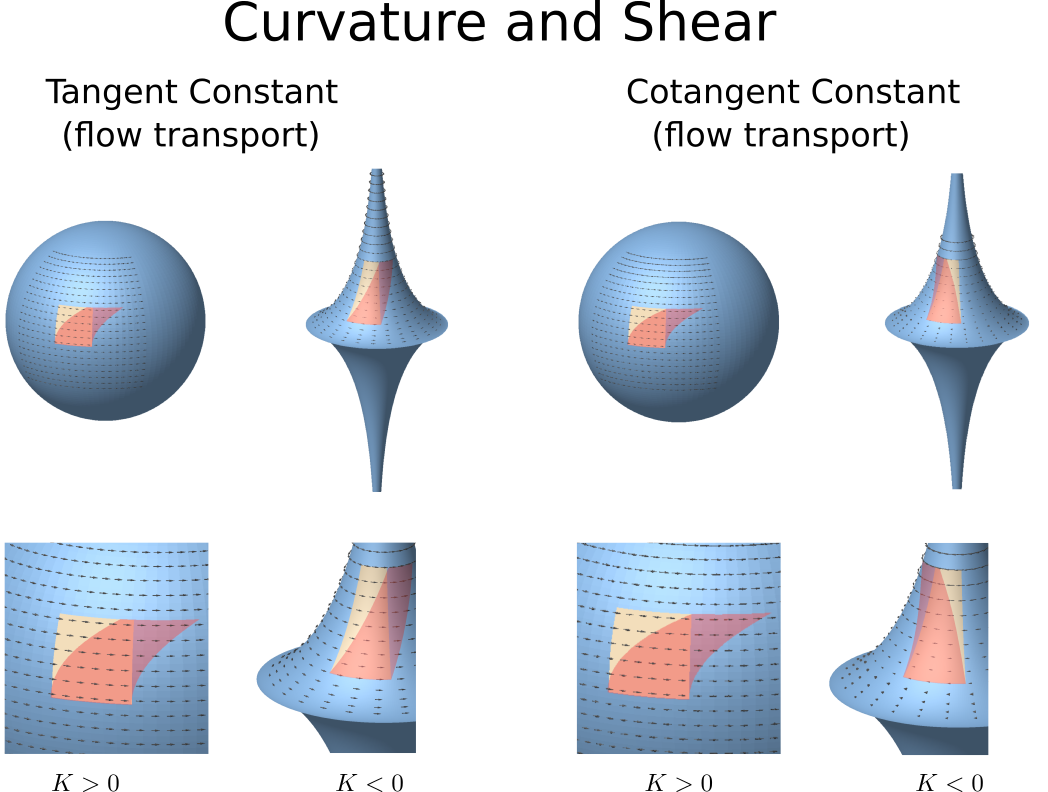}
\caption{Curvature and Shear.  In contrast to a flat membrane, material transported on a curved membrane can exhibit shear even by a flow having an effectively constant surface velocity.  We consider two surfaces (i) the sphere with constant Gaussian curvature $K > 0$ and (ii) the pseudosphere with constant Gaussian curvature $K < 0$.  For a rectangular patch of material on the surface (beige), we show how transport changes its shape over time (red).  On the left we show the transport for a velocity field with zero co-variant derivative $\nabla \mb{v} = 0$ (tangent vectors are constant).  On the right we show transport for a velocity field with zero dual exterior derivative $d\mb{v}^{\flat} = 0$ (co-tangent vectors are constant).  For either type of velocity field on the surface, in contrast to a flat surface, we see that the intrinsic curvature can result in shearing of the transported material.  This effect is captured in our hydrodynamic model by the Gaussian curvature term and exterior calculus in equations~\ref{equ_Stokes_geometric}. 
\label{fig:curvatureAndShear}}
\end{figure}

In contrast to a flat membrane, material transported on a curved membrane can undergo shear even by a flow having an effectively constant velocity field on the surface.  To investigate the role of intrinsic curvature of the surface, we consider flow on the sphere which has constant positive Gaussian curvature $K > 0$ and the pseudosphere which has constant negative Gaussian curvature $K < 0$~\cite{Pressley2001}, see Figure~\ref{fig:curvatureAndShear}.

For concreteness, we parametrize the sphere having Gaussian curvature $K = 1$ with the coordinates $(\theta,\phi)$ with $\mb{x} = \psi(\theta,\phi) = \left[\sin(\theta)\cos(\phi),\sin(\theta)\sin(\phi),\cos(\theta)\right]$.  We parametrize the pseudosphere having Gaussian curvature $K = -1$ with the coordinates
$(\theta,\phi)$ with $\mb{x} = \psi(\theta,\phi) = \left[\sech(\theta)\cos(\phi),\sech(\theta)\sin(\phi),\theta - \tanh(\theta)\right]$.

We first consider a flow having a velocity field $\mb{v}$ with zero co-variant derivative $\nabla \mb{v} = 0$ on the surface (constant tangent vector).  On both the sphere and pseudo-sphere a velocity having this property is given by $\mb{v} = \left[-\sin(\phi),\cos(\phi),0\right]$.   We remark that it is convenient here to express the velocity in terms of the $xyz$-components in $\mathbb{R}^3$  given by the embedding from the parametrization above.  For a curved surface, this provides the analogue to a flat surface of having a flow with constant velocity.  We find that the curvature results in shearing of the transported material.  Intuitively, this arises relative to the flat surface by the way 
intrinsic curvature requires distortion of the distance relationships between points on the surface.  More precisely, consider two points located at $(\theta_1,\phi_0)$ and $(\theta_2,\phi_0)$ with $\theta_2 > \theta_1 \geq 0$ in the upper hemisphere. While both points travel at exactly the same speed, the point $(\theta_2,\phi_0)$ which starts closer to the north pole will take less time to traverse fully around the $xy$-circular cross-section of the surface.  This curvature associated distortion of the distances results in shearing of the transported material.  This is illustrated in the panel on the left in Figure~\ref{fig:curvatureAndShear}.

We can also consider a flow having a velocity field $\mb{v}$ with dual field $\mb{v}^{\flat}$ having zero exterior derivative $d\mb{v}^{\flat} = 0$ (constant co-tangent vector field).  
The constant co-tangent case is motivated by the exterior calculus formulation of the fluid equations where for such an incompressible field the flow is determined only from the Gaussian curvature term, see equation~\ref{equ_Stokes_geometric}. 
We remark that while the co-tangent vector field $\mb{v}^{\flat} = v_b d\mb{x}^b$ is constant on both the sphere and pseudosphere, the velocity field $\mb{v} = v^a \partial_{x^a}$ on each surface is modulated by the local components of the inverse metric factor as $v^{a} = g^{ab} v_b$.  For any incompressible velocity field with zero exterior derivative $\mb{d}\mb{v}^{\flat} = 0$, according to equation~\ref{equ_Stokes_geometric} on any constant Gaussian curvature surface the force density $\mb{b}^{\flat}$ must also have zero exterior derivative $\mb{d}\mb{b}^{\flat} = 0$.  

To construct such a flow, we consider for the sphere $\mb{v}^{\flat} = -\mb{b}^{\flat}/2K = +d\phi$ and for the 
pseudosphere $\mb{v}^{\flat} = -\mb{b}^{\flat}/2K = -d\phi$.  The sign change in the fluid velocity arises from the way in which the Gaussian curvature effects the flow response to the force density, see equation~\ref{equ_Stokes_geometric}.  For the velocity field on the sphere, the inverse metric term $g^{\phi\phi} = 1/\cos^2(\theta)$ yields $\mb{v} = \left[-\sin(\phi)/\cos(\theta),\cos(\phi)/\cos(\theta),0\right]$.  For the pseudosphere, the inverse metric term $g^{\phi\phi} = 1/\sech^2(\theta)$ yields the velocity $\mb{v} = \left[\sin(\phi)/\sech(\theta),-\cos(\phi)/\sech(\theta),0\right]$.

We see that for both the sphere and pseudosphere the material transported under such a flow is sheared.  For the sphere and pseudosphere the shear is expected to be in the opposite direction arising from the difference in sign of the Gaussian curvature $K$ of the surface.  This is illustrated in the panel on the right in Figure~\ref{fig:curvatureAndShear}.  We remark that similar types of geometry and shear effects can be used for performing rheological experiments as was done in~\cite{Calvo1990}.

\section{Particle-Bilayer Coupling : Immersed Boundary Methods for Manifolds}
\label{sec:particleBilayerCoupling}

To model the motions of particles within the membrane, we compute a mobility tensor using approximations closely related to the \textit{Immersed Boundary Method} (IB)~\cite{Peskin2002,AtzbergerSELM2011,Atzberger2007a,Atzberger2006,Atzberger2007c,AtzbergerTabak2015}.  In IB the fluid-structure interactions are approximated by coupling operators that perform operations on the surrounding flow field to determine the particle velocity and perform operations yielding a force density field to account for particle forces~\cite{Peskin2002,AtzbergerSELM2011}.  We introduce IB approaches for manifolds to capture both the translational and rotational responses of inclusion particles to applied forces and torques when subject to coupling through the membranes hydrodynamics.  We show how our IB approaches can be used to compute an effective mobility tensor for these responses.

\subsection{Mobility Tensor}
\label{sec:mobilityTensor}
We express the mobility tensor $M$ for the velocity response of a collection of particles as
\begin{eqnarray}
\mb{V} = M \mb{F}.
\end{eqnarray}
The $\mb{V}$ is the collective vector of velocities and angular velocities of the particles and $\mb{F}$ is the collective vector of forces and torques applied to the particles.  For particle $i$, the velocity is given by 
$\mb{V}_i = \left[\mb{V} \right]_i$ and the particle force by 
$\mb{F}_i = \left[\mb{F} \right]_i$.  It is also convenient to decompose the mobility tensor into the components $M_{ij}$ where
\begin{eqnarray}
M = 
\left[
\begin{array}{llll}
M_{11} & M_{12} & \ldots & M_{1N} \\
\vdots & \vdots & \vdots & \vdots \\
M_{N1} & M_{N2} & \ldots & M_{NN} \\
\end{array}
\right].
\end{eqnarray}
The response of a single particle to a force applied directly to that particle is given by the diagonal block components $M_{ii}$.  The two-particle response associated with the velocity of particle $i$ in response to a force applied to particle $j$ is given by the off-diagonal block component $M_{ij}$.

\subsection{Coupling Operators $\Gamma$ and $\Lambda$ for Curved Surfaces}
\label{sec:mobilityOperators}

The mobility tensor for the interactions between the $i^{th}$ and $j^{th}$ particle is given by
\begin{eqnarray}
\label{def_mob_S}
M_{ij} = \Gamma_i \mathcal{S} \Lambda_j,
\end{eqnarray}
where we have the operators $\Gamma_i = \Gamma\left[\mb{X}^{i}\right]$ and $\Lambda_j = \Lambda\left[\mb{X}^{j}\right]$.  In the notation, we denote by $\mathcal{S}$ the solution operator for the hydrodynamic equations~\ref{equ_full_hydro_first}.  The velocity field for the hydrodynamics $\mb{v}(\mb{x})$ under the applied force density $\mb{f}(\mb{x})$ is given by $\mb{v} = \mathcal{S} \mb{f}$.  The operators $\Gamma,\Lambda$ approximate the fluid-structure interaction by modelling the velocity response and forces of the particles.  The force density generated by an applied force $\mb{F}$ on particle $j$ is given by $\mb{f} = \Lambda_j \mb{F}$.  In response, the velocity $\mb{V}$ of particle $i$ is given by $\mb{V} = \Gamma_i \mb{v}$.

\begin{figure}[H]
\centering
\includegraphics[width=8cm]{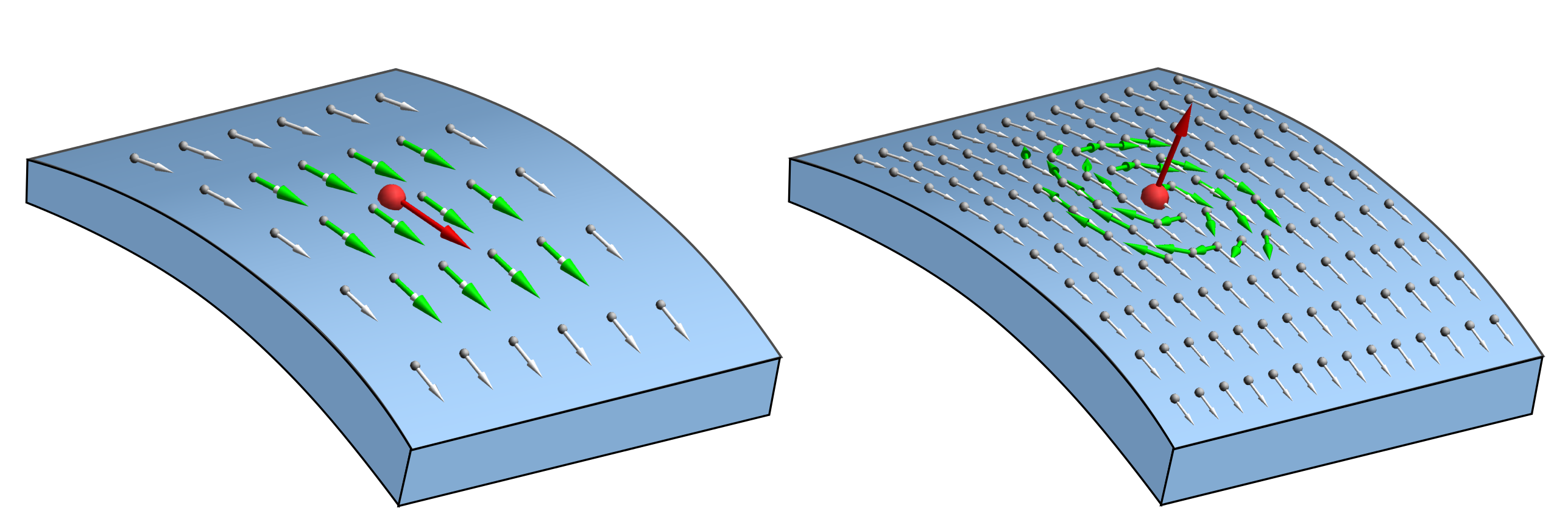}
\caption{For curved surfaces the coupling operators $\Gamma$ and $\Lambda$ must be consistent with the tangent bundle of the surface.  We use reference vector fields on the surface to construct the coupling operators $\Lambda$ and $\Gamma$.  We derive operators
$\Gamma$ and $\Lambda$ for translational and rotational motions
using the adjoint conditions in equations~\ref{equ_adjoint_cond} and ~\ref{equ_adjoint_discrete}.  On the left we show the reference vector field for translational responses (green).  On the right we show the reference vector field for rotational responses (green).   
\label{fig:gammaLambdaSurface}}
\end{figure}

Many choices can be made for the operators $\Gamma$ and $\Lambda$.  This can be used for both translational and rotational responses~\cite{AtzbergerSELM2011}.  To ensure that the approximate fluid-structure coupling is non-dissipative, it has been shown the operators must be adjoints~\cite{AtzbergerSELM2011,Peskin2002,AtzbergerTabak2015}.  We require for any choice of field $\mb{v}$ and vector $\mb{F}$ that the operators satisfy the adjoint condition
\begin{eqnarray}
\label{equ_adjoint_cond}
\langle \Gamma \mb{v} , \mb{F} \rangle & = & 
\langle \mb{v} , \Lambda \mb{F} \rangle.
\end{eqnarray}
The inner-products are defined as
\begin{eqnarray}
\langle \Gamma \mb{v} , \mb{F} \rangle & = & 
\sum_i \left[ \Gamma \mb{v} \right]_i \cdot \left[\mb{F}\right]_i \\
\langle \mb{v} , \Lambda \mb{F} \rangle & = & \int_{\Omega} 
\mb{v}(\mb{x})\cdot \left(\Lambda \mb{F}\right)(\mb{x}) d\mb{x}
\end{eqnarray}
where $\cdot$ denotes the dot-procuct in the embedding space $\mathbb{R}^3$.
We use the notation $\Gamma^T = \Lambda$ to denote succinctly the adjoint condition~\ref{equ_adjoint_cond}.  We remark that from equation~\ref{def_mob_S} this condition has the desirable consequence of yielding a symmetric mobility tensor $M$ when $\mathcal{S}$ is symmetric.  

To obtain the translation and rotational responses of the particles, we introduce the operators
\begin{eqnarray}
\Gamma \mb{v} & = & \int_{\Omega} \mb{W}\left[\mb{v}\right] (\mb{y}) d\mb{y} \\
\Lambda \mb{F} & = & \mb{W}^*\left[ \mb{F} \right](\mb{x}).
\end{eqnarray}
The $\mb{X}$ denotes the collective vector of particle locations.  The $i^{th}$ particle is at location $\left[\mb{X}\right]_i$.

To obtain the particle velocity in response to the hydrodynamic flow, the tensor $\mb{W}$ serves to average by sampling and weighing the velocity values on the surface.  For a particle force the adjoint tensor $\mb{W}^*$ serves to produce a force density field. 

For a curved surface, $\mb{W}$ must be chosen carefully.  A simple form which is widely used in IB is to use a scalar weight $\mb{W}^*[\mb{F}] = \eta(\mb{y} -\mb{X}) \mb{F}$ where $\eta$ is the Peskin $\delta$-function~\cite{AtzbergerSELM2011,Peskin2002}.  However, for a curved surface this is not a good choice since the force density field produced by $\Lambda\mb{F} = \eta(\mb{y} -\mb{X}) \mb{F}$ is not in the tangent space of a curved surface.  Similarly, the averaging procedure $\Gamma$ may produce a particle velocity which is not in the tangent space.

For curved surfaces, we use a more geometrically motivated operator of the form 
\begin{eqnarray}
\mb{W}\left[\mb{v}\right] & = & 
\sum_i \mb{w}^{[i]}\left[\mb{v}\right](\mb{x})   \\
& = & \sum_i  \mb{w}^{[i],\alpha}\left[\mb{v}\right]  \partial_{\alpha}|_{\mb{X}^{[i]}}  \\
\nonumber
 & = & \sum_i
 \left(\left(w^{[i]}\right)^{\alpha}_{\beta}v^{\beta}\right)  \partial_{\alpha}|_{\mb{X}^{[i]}}.
\end{eqnarray}
The sum $i$ runs over the particle indices and the $\partial_{\alpha}|_{\mb{X}^{[i]}}$ denotes the tangent basis vector in direction $\alpha$ at location $\mb{X}^{[i]}$.  The square brackets $[i]$ are used to help distinguish entries not involved in the Einstein conventions of summation.  This can be interpreted as the procedure of obtaining the average velocity for particle $i$ by using for each coordinate direction $\alpha$ the inner-product of the velocity field $\mb{v}$ with the reference vector field $\mb{q}^{\alpha} = \left(\mb{w}^{[i],\alpha}\right)^{\sharp} = \left(w^{[i],\alpha}\right)^{\gamma}\partial_{\gamma}$. 
The adjoint tensor yielding the local force density is given by 
\begin{eqnarray}
\mb{W}^*\left[\mb{F}\right] & = & \sum_i 
\left(\mb{w}^{[i],\alpha}\right)^{\sharp} F^{\alpha}  \\
& = & \sum_i 
\left(w^{[i],\alpha}\right)^{\gamma} F^{\alpha}      \partial_{\gamma}.
\end{eqnarray}

For translational motions we use the reference vector fields of the form $\mb{q}^{\theta} = \psi(\mb{x} - \mb{X}^{[i]}) \partial_{\theta}$ 
and $\mb{q}^{\phi} = \psi(\mb{x} - \mb{X}^{[i]})/\cos(\theta) \partial_{\phi}$, where $\psi(r) = C \exp(-r^2/2\sigma^2)$.  
For rotational responses we use the reference vector field on the surface
$\mb{q}^{n} = \psi(\mb{x} - \mb{X}^{[i]})\left( \mb{n} \times (\mb{x} - \mb{X}^{[i]})\right)$.  We emphasize that we only use these expressions for a coordinate chart chosen so that the particle location $\mb{X}^{[i]}$ is away from a polar singularity, see Appendix~\ref{sec:coord_charts}.  The reference velocity fields are illustrated in Figure~\ref{fig:gammaLambdaSurface}.

In practice for the spatially discretized system, the operators and the associated fields they generate can be expressed conveniently as 
\begin{eqnarray}
V^{i} & = & \Gamma_{m}^{i} v^{m} \\
f^m & = & \Lambda^{m}_j {F}^{j}.
\end{eqnarray}
The index $m$ corresponds to the discrete degrees of freedom, such as the index of a lattice point or harmonic mode, and the $i$ and $j$ index the components of the vector.  We have 
\begin{eqnarray}
V^k F^k = \Gamma_m^{k} v^m {F}^{k} = v^m \Lambda^m_{k} {F}^{k} = v^m f^m.
\end{eqnarray}
The adjoint condition can be expressed as
\begin{eqnarray}
\label{equ_adjoint_discrete}
\Gamma_m^{k} = \Lambda_{k}^{m}.
\end{eqnarray}
In practice, we define the operator $\Lambda$ in numerical calculations using the specified reference velocity fields $\mb{q}^{\alpha}$ above to generate the force density at lattice sites on the sphere surface.  Using the sparse matrix representation of this operation for $\Lambda$, equation~\ref{equ_adjoint_discrete} provides the adjoint velocity averaging operator $\Gamma$.  This approach for developing consistent operators $\Lambda$ and $\Gamma$ on the sphere also extends straight-forwardly to immersed boundary approximations on more general curved surfaces and manifolds.

\section{Dynamics of Inclusion Particles Embedded in Spherical Bilayers}
\label{sec:particleMobilities}
\begin{figure}[H]
\centering
\includegraphics[width=8cm]{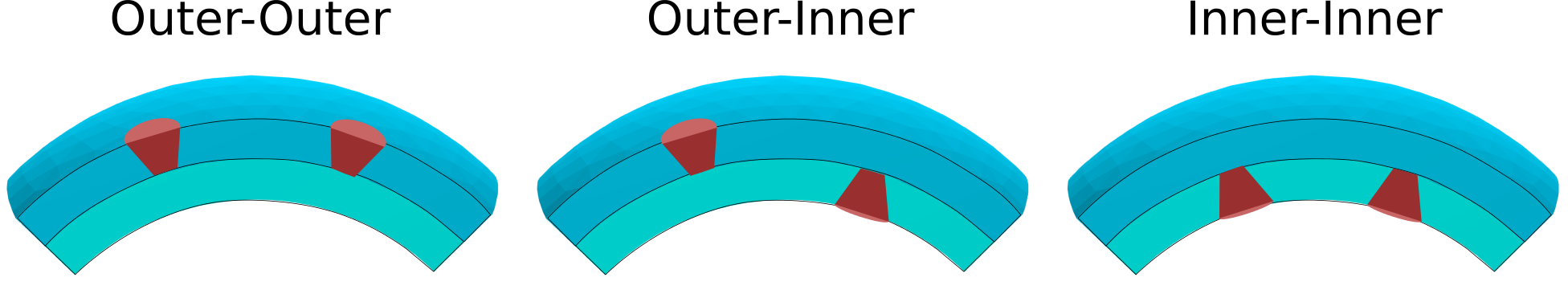}
\caption{We consider the case when each embedded inclusion particle only spans one of the bilayer leaflets.  We consider the interaction cases when particles are both in the same leaflet or are in different leaflets. \label{fig:inclusionSchematic}}
\end{figure}

For particle inclusions embedded within spherical lipid bilayer membranes, we investigate their translational and rotational motions in response to applied forces and torques.  We consider the case when each embedded inclusion particle only spans one of the fluid bilayer leaflets, see Figure~\ref{fig:inclusionSchematic}.  We investigate the mobility of these inclusions when varying the (i) vesicle radius, (ii) membrane viscosity, (iii) solvent viscosity, and (iv) intermonolayer slip.  We investigate the four interaction cases (i) outer-outer, (ii) outer-inner, (iii) inner-outer, and (iv) inner-inner.  We also investigate the coupled motions for the four cases (i) translation-translation, (ii) translation-rotation, (iii) rotation-translation, and (iv) rotation-rotation.

We express the translational and rotational responses as
\begin{eqnarray}
\left[
\begin{array}{l}
\mb{V} \\
\boldsymbol{\omega} 
\end{array}
\right]
=
\mb{M}
\left[
\begin{array}{l}
\mb{F} \\
\boldsymbol{\tau}
\end{array}
\right]
\end{eqnarray}
where we decompose the mobility tensor into the blocks
\begin{eqnarray}
\mb{M} = 
\left[
\begin{array}{ll}
M_{tt} & M_{tr} \\
M_{tr} & M_{rr} \\
\end{array}
\right].
\end{eqnarray}
In the notation, the $\mb{V}$ denotes the collective translational velocities and $\boldsymbol{\omega}$ the collective rotational angular velocities.  The $\mb{F}$ denotes the collective forces applied to particles within the inner and outer leaflets. The $\boldsymbol{\tau}$ denotes the collective torques applied to particles within the inner and outer leaflets.

We denote the different ways in which the forces $\mb{F}$ and torques $\boldsymbol{\tau}$ couple to the particle translational motions $\mb{V}$ and rotational motions $\boldsymbol{\omega}$ using the notation $M_{XY}$.  The $X$ denotes the response as either translation (t) or rotational (r).  The $Y$ denotes the type of applied force as either standard force (t) or torque (r).  The mobility components can be further decomposed into $M_{XY,i_1,\ell_1,i_2,\ell_2}$ where $i_k$ denotes the location of the $i_k^{th}$ particle. The leaflets in which the inclusions are embedded is denoted by $\ell_k \in \{\mbox{inner}, \mbox{outer}\}$.

There are a few notable differences between spherical fluid membranes and flat fluid membranes.  In the flat case, the membrane domain is often treated as effectively infinite and for theoretical convenience often as having periodic boundary conditions.  In the spherical case, the membrane is intrinsically of finite area.  Also for a sphere, as consequence of the topology, any in-plane hydrodynamic flow must have a singularity~\cite{Jarvis2004}.  For the solvent fluid, flat membranes have fluid extending over an infinite domain symmetrically on both sides. 
In the spherical case, this symmetric is broken with solvent fluid trapped within the interior in a region of finite volume and with solvent fluid extending exterior over an infinite domain.  The curvature of the membrane surface can also play an important role in the hydrodynamics.  This is particularly apparent from the Gaussian term that appears in equation~\ref{equ_full_hydro_first} and the effects we discussed in Section~\ref{sec:curvatureAndShear}.

We investigate the mobility of inclusion particles within spherical bilayers in a few different regimes.  We consider the characteristic scales introduced in Section~\ref{sec:charScales}.  The regime with $\LRratio = \LRratioExp \gg 1$ and for $\gammaMuRatio = \gammaMuRatioExp$ with $\gammaMuRatio^{-1} \LRratio = 1$ corresponds to the case when the hydrodynamic flow is dominated by the intramembrane viscosity and intermonolayer slip.  In this regime, for a force density having a non-zero net torque, the flow is approximated well by the leading order spherical harmonic modes with $\ell = 1$, see equation~\ref{equ_Stokes_SPH_sol2_defA_ells_nonDim}.  The intramembrane viscosity strongly couples the surface fluid resulting in a flow that is a rigid body rotation of the entire spherical shell, see Figure~\ref{fig:singleParticleFlowResponse}.  Parameter values are given in Table~\ref{table:defaultParams}.

Mathematically, this arises from the dominant spherical harmonic modes with degree index $\ell = 1$ and order index $m = -1,0,1$. Using the exterior calculus formation we apply the generalized $\mbox{curl} = -\star d$ on the surface to the vector potential $\Phi$ given by a linear combination of the harmonic modes of degree $\ell = 1$.  This yields for the velocity field on the surface that of a rigid body rotation, see equation~\ref{equ_velFieldRep}.  In the case when the surface force has zero net torque in the regime $\LRratio \gg 1$, $\gammaMuRatio^{-1} \LRratio = 1$, the leading order flow is determined by the intramembrane viscosity and intermonolayer slip and depends on the higher-order moments of the torque of degree $\ell > 1$.

The regime with $\LRratio \ll 1$ and $\gammaMuRatio \ll 1$ corresponds to the case when the traction stress from the entrained external solvent fluid dominates the hydrodynamic response relative to the intramembrane viscosity and intermonolayer slip.  This results in more localized flow within the surface, see Figure~\ref{fig:singleParticleFlowResponse}.  

We remark that the regime when $\gammaMuRatio \gg 1$ corresponds to the case when the intermonolayer slip strongly couples the hydrodynamic flow between the two leaflets to make them nearly identical.  This effectively doubles the intramembrane viscosity.  

We have presented a few regimes indicating the contributing factors in the hydrodynamic responses and the interplay between the entrained solvent fluid, intramembrane viscosity, and intermonolayer slip.  We now discuss some features of the hydrodynamic response that arise from the geometry of the spherical membrane.

\multicolinterrupt{
\begin{figure}[H]
\centering
\includegraphics[width=16cm]{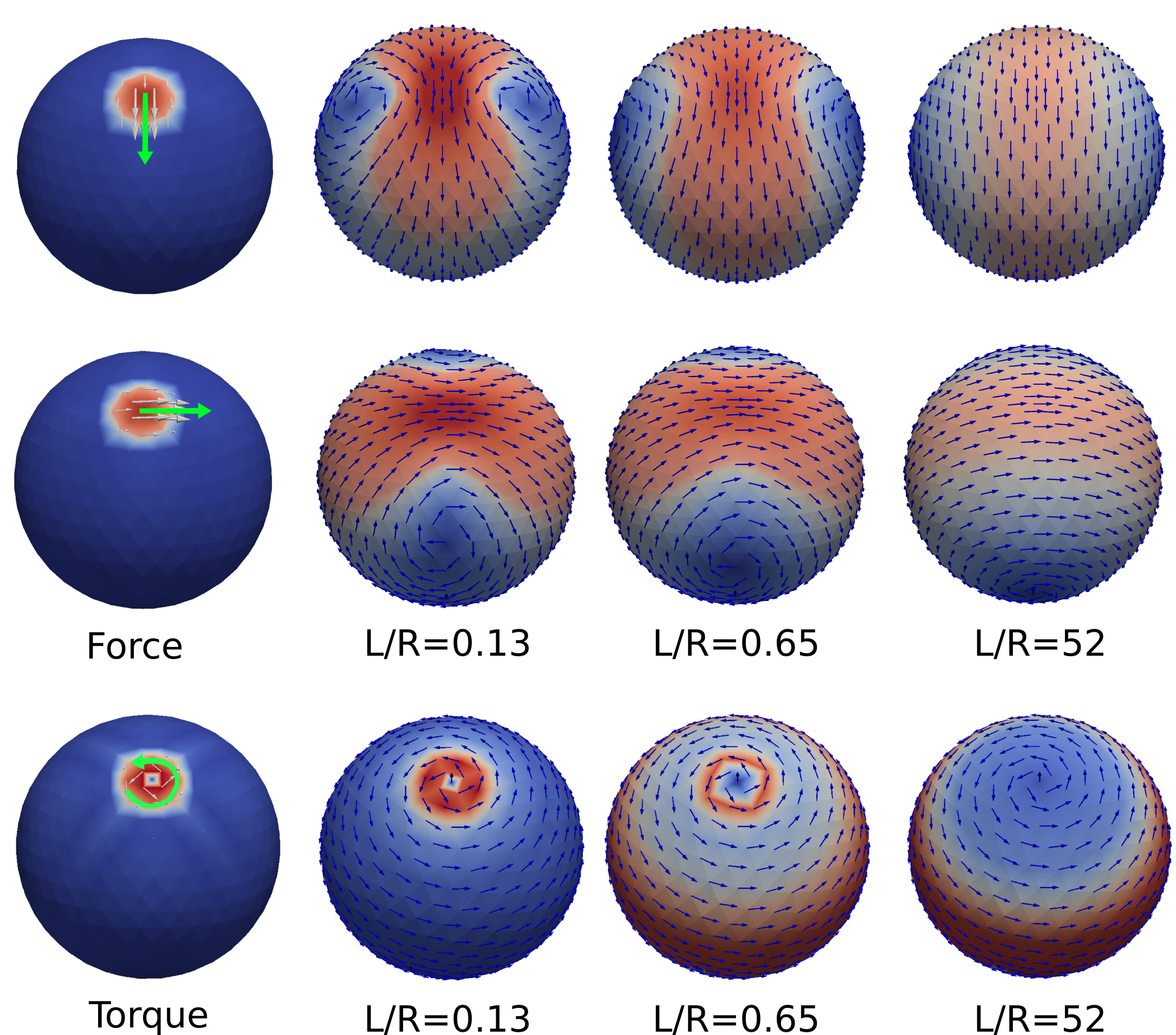}
\caption{Hydrodynamic flow in response to force acting on an  inclusion particle.  The $L/R = \LRratio$ is the relative Saffman-Delbr\"uck length-scale scaled by the vesicle radius.  For small intramembrane viscosity the force produces a localized hydrodynamic flow on the surface.  As the membrane viscosity increases the hydrodynamic flow becomes less localized and eventually approaches the velocity field of a rigid body rotation of the sphere.  The flow exhibits two vortices with locations that migrate toward the equatorial poles as the viscosity increases.  Parameter values in Table~\ref{table:defaultParams}.  
\label{fig:singleParticleFlowResponse}}
\end{figure}
}

\subsection{Vortices and Membrane Viscosity}
\label{sec:vortices}

\begin{figure}[H]
\centering
\includegraphics[width=8cm]{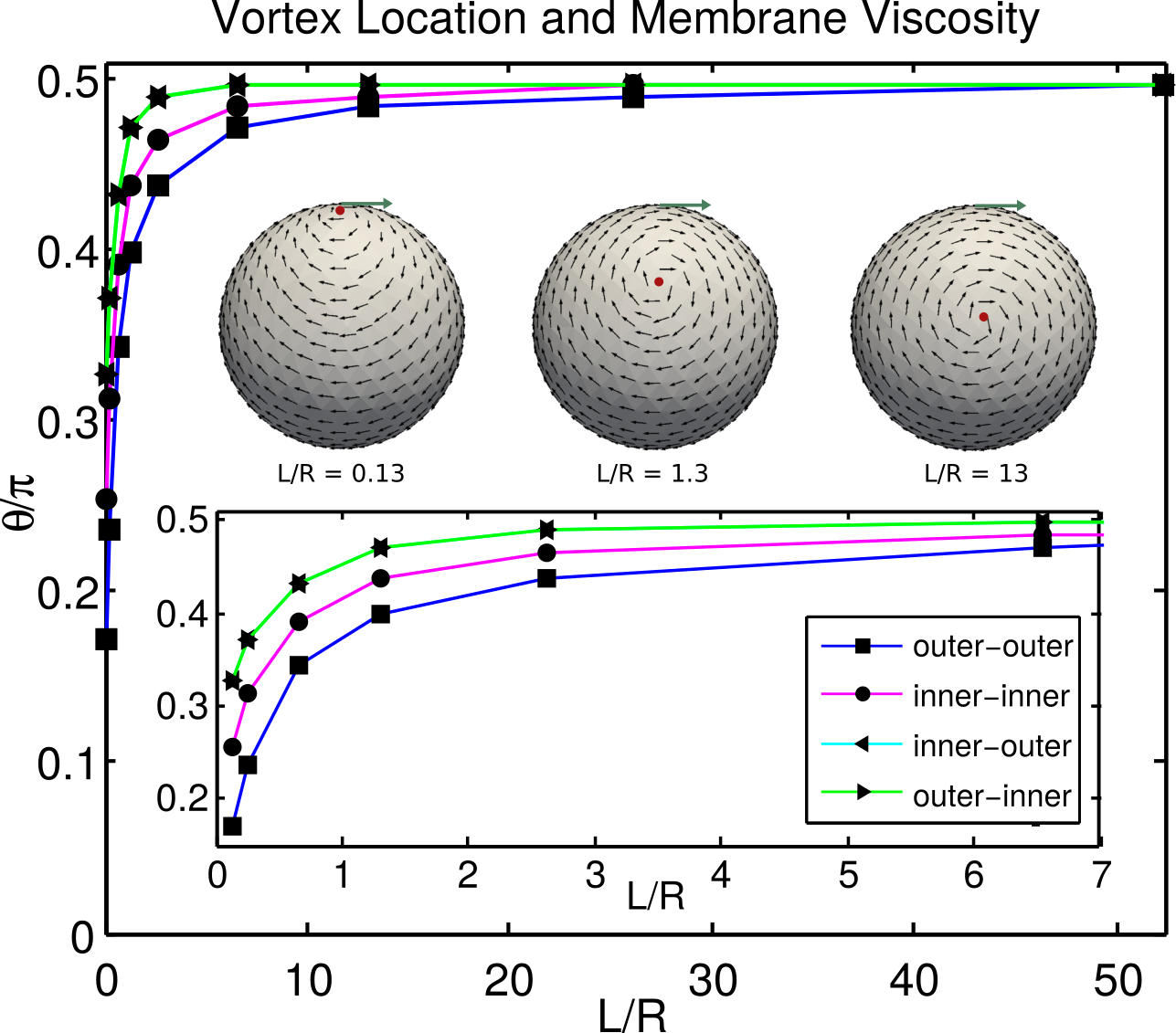}
\caption{Vortex Location and Membrane Viscosity.  For a force applied to a particle in the outer leaflet located at the north pole, we show as the shear viscosity is varied how the vortex location changes in the outer and inner leaflets.  In the nomenclature X-Y in the figure caption, X refers to the leaflet of the applied force and Y the leaflet of the flow response.  For low viscosity the vortices are near the north pole $\theta = 0$.  As viscosity increases the vortices migrate toward the equator $\theta = \pi/2$.  The inset left to right shows typical progression in flows of the vortex location.  The intermonolayer slip $\gammaMuRatio = \gamma/\gamma_0 = 4$ moderately couples the inner leaflet to the outer leaflet.  We find this results in a flow within the inner leaflet with a vortex location closer toward the equator.  Parameter values in Table~\ref{table:defaultParams}.  
 \label{fig:vortexLocation}}
\end{figure}

As a consequence of the spherical topology of the membrane, any hydrodynamic flow within the surface must contain a singularity~\cite{Jarvis2004}.  We consider the case of an inclusion particle located at the north pole of the sphere and subjected to a force.  These singularities manifest in the flow as two vortices of opposite sign, see Figures~\ref{fig:singleParticleFlowResponse}.  The location of these vortices depends on $\LRratio = L/R$ characterizing the relative strength of the intermembrane shear viscosity vs the solvent traction stress.  For small $\LRratio$ the vortices start near the north pole and as $\LRratio$ increases they migrate toward the equator, see Figures~\ref{fig:singleParticleFlowResponse}.  For a force applied to a particle in either the outer leaflet or inner leaflet we consider how the vortex location changes as the viscosity of the membrane is varied.  We show the vortex locations in the outer and inner leaflets in Figure~\ref{fig:vortexLocation}.  In this case we vary $\LRratio = L/R$ and keep fixed $\gammaMuRatio = \gamma/\gamma_0$ where $\gamma_0 = \mu_f/R$ with parameters in Table~\ref{table:defaultParams}.  We remark that these results can be used as a reference to estimate the membrane shear viscosity by making observations of the vortex locations of the fluid flow within the leaflets.  Some recent experimental work to estimate the membrane viscosity of vesicles using vortex locations can be found  in~\cite{Woodhouse2012,HonerkampSmith2013,Dimova1999}.

\subsection{Self-Mobility and Coupled-Mobility}
\label{sec:mobilityResults}
We next consider the hydrodynamic responses when a force or torque is applied to an inclusion particle embedded in the outer leaflet when the center of the sphere is held fixed.  We take as our convention that this particle is embedded at a pole where we parametrize the sphere with $(\theta,\phi) = 0$.  We then consider 
how the resulting hydrodynamic flows within the inner or outer leaflets within the spherical bilayer couple the translational motions and rotational motions of inclusion particles at other locations.  Throughout, we use the base-line parameters given in Table ~\ref{table:defaultParams}.  These parameters correspond to the non-dimensional regime with $\LRratio = 0.65$ and $\gammaMuRatio = 4.0$.
\begin{table}[H]
\footnotesize
\centering
\begin{tabular}{|l|l||l|l|}
\hline
\rowcolor{LightGrey}
Parameter & Value & Parameter & Value \\
\hline
$R_{-}$ & 14 $\sigma$ & $\mu^{\pm} = \mu_f$ & 383 $m_0/\tau\sigma$ \\  
\hline
$R_{+}$ & 16.6 $\sigma$ & $\mu_m$ & 3830 $m_0/\tau$ \\  
\hline
$R$    & 15.3 $\sigma$ & $\gamma$ & 100 $m_0/\tau\sigma^2$  \\ 
\hline
$\sigma$ & 1 nm & $m_0$ & 1 amu \\ 
\hline
$\tau$ & 0.64 ps & $\epsilon$ & 2.5 amu$\cdot$nm$^2$/ps$^2$ \\ 
\hline 
\end{tabular}
\caption{Vesicle Parameters.  We use these default parameters throughout our discussions unless specified otherwise.  These parameters correspond to the non-dimensional regime with $\LRratio = L/R = 0.65$ and $\gammaMuRatio = \gamma R/\mu_f = 4.0$.  
\label{table:defaultParams}}
\end{table}

\begin{figure}[H]
\centering
\includegraphics[width=8cm]{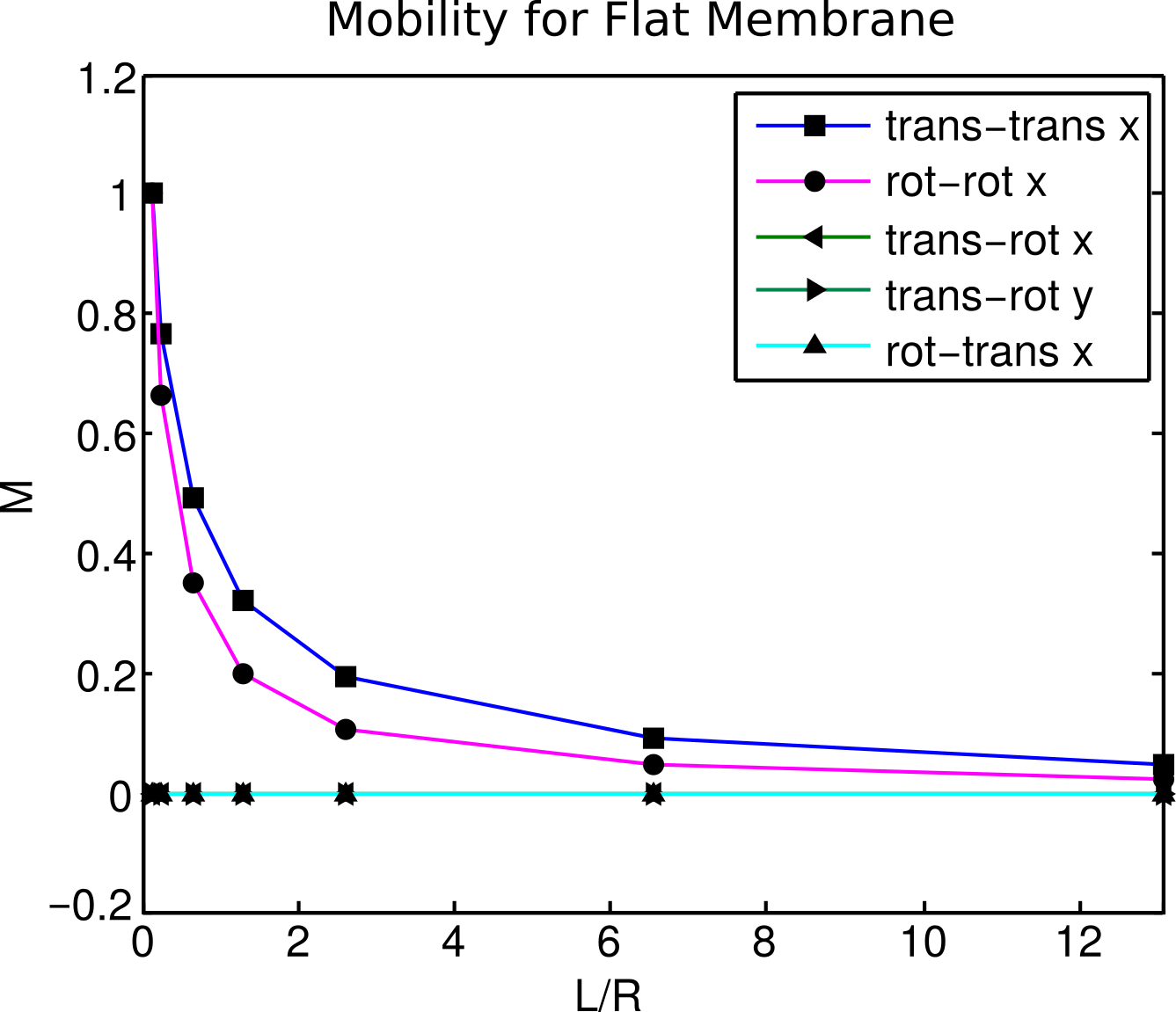}
\caption{Mobility for Flat Membranes.  For a force or torque applied to a particle embedded within a large flat membrane, we show as the intramembrane viscosity is varied the translational and rotational responses of inclusion particles.  Normalized by the mobility response when $\LRratio = L/R = 0.13$.  These results were obtained from solving in Fourier space the hydrodynamic flow in response to an applied force density following closely the analytic approach presented in~\cite{Saffman1976, Oppenheimer2009} and our method for computing the mobility tensor discussed in Section~\ref{sec:mobilityOperators}. 
 \label{fig:flatMembrane}}
\end{figure}

We investigate the roles played by the bulk solvent fluid, the intramembrane viscosity, and the intermonolayer slip.  We use our methods to compute profiles of the mobility responses at different locations when varying the intramembrane viscosity and intermonolayer slip in Figure~\ref{fig:singleMobilityProfiles}.  We show how the mobility varies when changing the intermonolayer slip and membrane viscosity in Figures~\ref{fig:viscAll} and~\ref{fig:slipAll}.  For comparison we also compute the mobility responses within a flat membrane shown in Figure~\ref{fig:flatMembrane}.

Before discussing in more detail these results, we make a few remarks concerning how the mobility results are reported.  The responses are shown along the two great circles on the sphere corresponding to the intersection with the $xy$-plane and the $xz$-plane.

\begin{figure}[H]
\centering
\includegraphics[width=8cm]{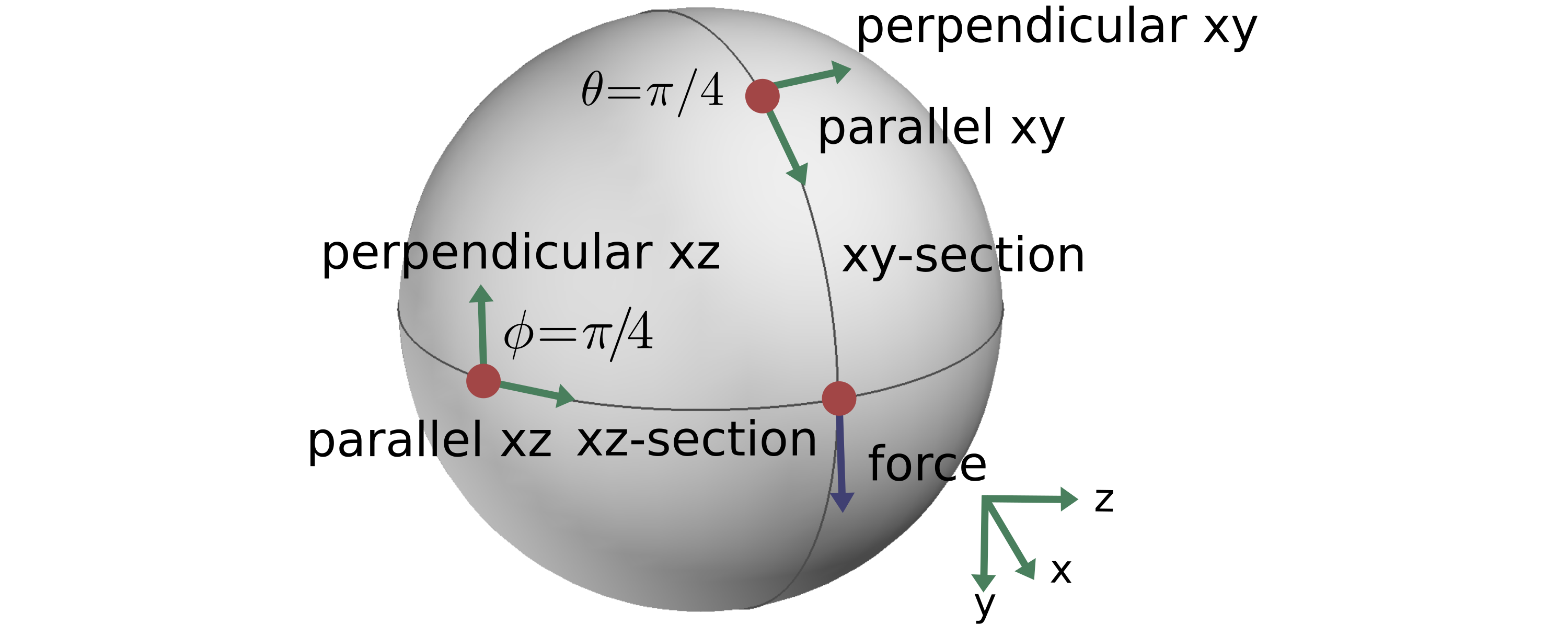}
\caption{Cross Sections of the Sphere and Conventions.  We consider the hydrodynamic responses when a force or torque is applied to an inclusion particle.  For convenience in our calculations, we use by convention the coordinates for the inclusion particle $\mb{X} = (x,y,z) = (1,0,0)$ and we apply force to the inclusion particle in the direction $\mb{F} = (f_x,f_y,f_z) = (0,1,0)$.  To characterize the hydrodynamic responses, we consider the cross-sections of the sphere in the $xy$-plane and the $xz$-plane.  This gives two great circles of the sphere.  We consider the velocity in the directions parallel and perpendicular to the tangents of each of the respective great circles.  
 \label{fig:diagramCrossSections}}
\end{figure}

We consider the velocity responses in the parallel $\parallel$ and perpendicular $\perp$ directions along each of these curves.  We normalize all of the mobility results by comparing to the case of large intramembrane viscosity $\LRratio = L/R = 48$ for the leaflet or large intermonolayer slip corresponding to $\gammaMuRatio = \gamma R/\mu_f = 32$.  This regime provides a reference case corresponding to the situation when the large intramembrane viscosity yields an effective rigid body rotation of the spherical shell within the bulk solvent fluid or when the two leaflets are tightly coupled.

We remark that this is in contrast to the flat membrane case where the mobility tends to zero as $\LRratio = L/R$ becomes large.  In the flat membrane case, we normalize instead our reported results by the self-mobility when $\LRratio = L/R = 0.1$.  Given the mobility model for the hydrodynamic responses discussed in Section~\ref{sec:mobilityTensor}, the self-mobility on the sphere for each type of coupling is given in our model by the results reported at location $(\theta, \phi) = 0$.

The mobility profiles reveal a number of interesting aspects of the hydrodynamic coupling between inclusion particles and leaflets.  We find that the intermonolayer slip and curvature yield coupling for particles embedded in the inner leaflet significantly different than for particles embedded in the outer leaflet.  For a force or torque applied to a particle embedded in the outer leaflet, the intermonolayer slip yields a flow for the inner leaflet with recirculation over a larger scale.  This is seen when looking at the vortex locations when applying force at the north pole, where the intermonolayer slip plays a role pushing the vortex location of the inner leaflet closer to the equatorial poles, see Figure~\ref{fig:vortexLocation}.  

We see this can result in both the translational motions and rotational motions of a particle within the inner leaflet moving in the opposite direction of an inclusion particle within the outer leaflet at the same location.  This is seen for the smallest viscosities and intermonolayer slips for the translation-translation responses at location $xz$ with $\phi = \pi/4$ and for the rotation-rotation responses at location $xy$ with $\theta = \pi/4$, see Figure~\ref{fig:viscAll} and~\ref{fig:slipAll}.  

\newpage

\multicolinterrupt{
\begin{figure}[H]
\centering
\includegraphics[width=16cm]{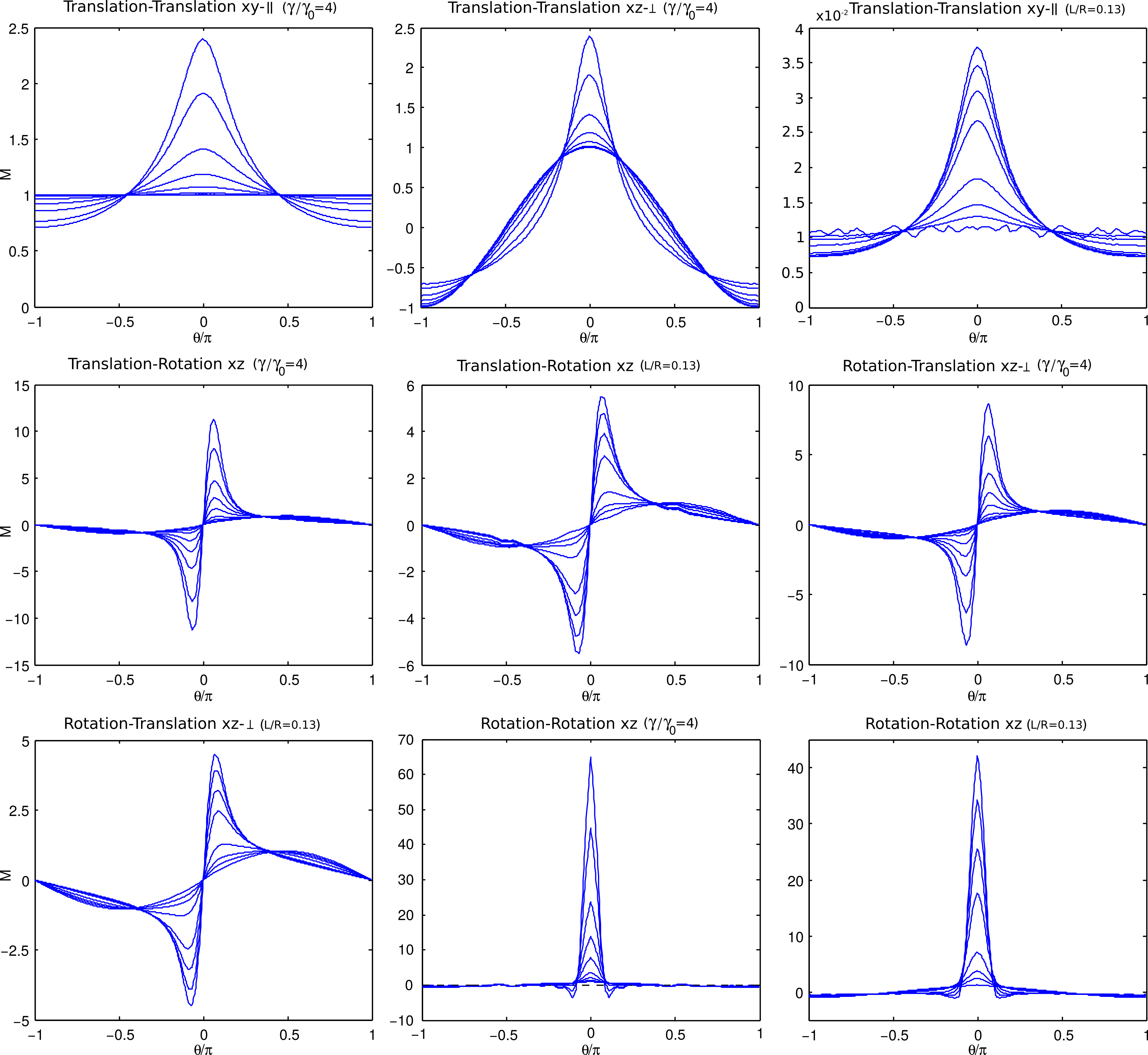}
\caption{Mobility profiles of inclusion particles when varying the
membrane viscosity and intermonolayer slip.  In each case a force or a torque is applied to a single inclusion particle within the outer leaflet located at $(\theta,\phi) = 0$.  The resulting inclusion particle hydrodynamic response within the outer leaflet or inner leaflet is shown in terms of the mobility $M$ defined in Section~\ref{sec:mobilityTensor}.  We use in the nomenclature in the titles of X-Y to indicate a forcing of type X and a response of type Y.  We normalize the mobility by the self-mobility response obtained in the case when $\LRratio = L/R = 48$ and $\gammaMuRatio = \gamma R/\mu_f = \gamma/\gamma_0 = 32$.  The intramembrane viscosity or intermonolayer slip is held fixed in panels displaying respectively $\LRratio = L/R = 0.13$ or $\gammaMuRatio = \gamma/\gamma_0 = 4$.  All figures show the outer leaflet response with the exception of the figure on the upper-right for the translation-translation response which shows how the inner leaflet responds to increasing intermonolayer slip.  The other panels show the dependence of the mobility response of inclusion particles embedded within the outer-leaflet when increasing the membrane viscosity as $\LRratio = L/R = 0.13, 0.26, 0.65, 1.3, 2.6, 6.5, 13, 26, 52$.
The curve with largest amplitude at $\theta = 0$ corresponds to the largest local mobility response which occurs for the smallest membrane viscosity.  The panels show the dependence of the mobility response of inclusion particles embedded within the inner-leaflet when increasing the intermonolayer slip as $\gammaMuRatio = \gamma/\gamma_0 = 0.040, 0.10, 0.40, 1.0, 4.0, 8.0, 16, 32$.
The curve with smallest amplitude at $\theta = 0$ shows the smallest mobility response corresponds in each case to the smallest intermonolayer slip. 
\label{fig:singleMobilityProfiles}}
\end{figure}
}

\newpage

For the translational and rotational response to forces in the outer leaflet, we find that the intermonolayer coupling smooths the flow over a larger scale within the inner leaflet.  

We next consider for fixed intramembrane viscosity how the intermonolayer slip effects the flow.  We see for a force acting on the outer leaflet as the intermonolayer slip becomes small the flow within the inner leaflet approaches a rigid body rotation, see the bottom curve in the upper-right panel of Figure~\ref{fig:singleParticleFlowResponse}.  

From an analysis of the hydrodynamic response equation~\ref{equ_Stokes_SPH_sol2_defA_ells}, we have two interesting cases for the modes of the inner leaflet: (i) $\ell = 1$ and (ii) $\ell > 1$.  In the first case, the inner-leaflet rotates as a rigid spherical shell and entrains the fluid trapped within to a rigid body motion.  As a consequence there is no traction stress with the external solvent fluid for the inner leaflet and no intramembrane shear stress.  This means there are no other stresses acting against the intermonolayer drag so $-\gamma(a_{\ell}^{-} - a_{\ell}^{+}) = 0$ and the inner leaflet matches the outer leaflets rotation with $a_{\ell}^{-} = a_{\ell}^{+}$ for $\ell = 1$.  

In the second case with $\ell > 1$, the intramembrane stress and traction stress balance the intermonolayer drag.  In this case, the hydrodynamic modes of the inner leaflet scale in proportion to the intermonolayer slip and the modes of the outer leaflet.  As the intermonolayer slip decreases, the modes $a_{\ell}^{-}$ of the inner leaflet become small for $\ell > 1$. 

This can be seen mathematically from equation~\ref{equ_Stokes_SPH_sol2_defA_ells} where the inner leaflet modes satisfy  $a_{\ell}^{-} = -\left(\left(A_2^{\ell}/\gamma\right) - 1 \right)^{-1} a_{\ell}^{+}$.  This can be expressed as 
\begin{eqnarray}
\nonumber
a_{\ell}^{-} = -\Pi_3\left(2 - \ell(\ell + 1) - \Pi_1^{-1}(\ell - 1) - \Pi_3 \right)^{-1} a_{\ell}^{+}
\end{eqnarray}
where for convenience we denote $\Pi_3 = \Pi_2^{-}/\Pi_1^{-}$.  For $\ell = 1$ this shows $a_{\ell}^{-} = a_{\ell}^{+}$ independent of the magnitude of $\Pi_3 \neq 0$.  For $\ell > 1$, we have as the intermonolayer slip becomes small $\Pi_3 \ll 1$ the hydrodynamic response for the mode of the inner leaflet with $\ell > 1$ become small $a_{\ell}^{-} \ll 1$.  This shows that the resulting hydrodynamic responses in the inner leaflet become dominated by the rigid rotation mode $\ell = 1$ for small intermonolayer slip.  This can be seen in the upper-right panel of Figure~\ref{fig:singleMobilityProfiles}.

This has a number of interesting consequences for the motions of inclusion particles embedded within leaflets of spherical bilayers.  From the different hydrodynamics of the two spherical shells, we have that for small intermonolayer slip the self-mobility and coupled-mobilities can result in large motions when forces or torques act on an inclusion particle within the inner leaflet.  For small intermonolayer slip this arises since the rigid body mode $\ell = 1$ of the hydrodynamic response for the inner leaflet is not damped by the trapped solvent fluid but only by the weak intermonolayer coupling.  This manifests in a near rigid rotation of the inner leaflet and a large self-mobility and coupled-mobility in response to an applied force or torque, see Figure~\ref{fig:slipAll}.  

We remark that it is important to keep in mind this behaviour arises when forces applied to inclusion particles result for the inner leaflet in a force moment with non-zero net torque.  This is what drives a significant hydrodynamic response for the rotation mode $\ell = 1$.  In contrast, for the case of a collection of inclusion particles with total force acting on the inner leaflet that yields a zero net torque, super-position of the particle hydrodynamic responses cancel for the $\ell = 1$ mode and the behaviour of large motions for inclusions from the rigid shell rotation is suppressed.  This means for inclusion particles embedded within spherical bilayers it is important to consider carefully the different ways forces and net torque act on the leaflets.

As the intermonolayer slip becomes large, the hydrodynamic flows within each of the two leaflets approach a common velocity.  The self-mobility of inclusion particles embedded in the inner and outer leaflet also approach a common value.  It is interesting to note that the common value is not simply $1/2$ of the self-mobility for the uncoupled leaflets, see Figure~\ref{fig:slipAll}.  This arises from the asymmetric way in which the leaflets couple to the external solvent.  For the outer leaflet the solvent is within an unbounded domain exterior to the spherical shell.  For the inner leaflet the solvent is within a bounded domain trapped interior to the spherical shell.  As a consequence, we see there are different tractions acting on the inner and outer leaflet, see equation~\ref{equ_Stokes_SPH_sol2_defA_ells}.  As we saw for the rigid rotation mode $\ell = 1$, there is no traction stress on the inner leaflet since the solvent fluid rigidly rotates within the spherical shell but there is traction stress from the solvent on the outer leaflet.  For the other modes $\ell > 1$ there continue to be asymmetries in the strength of the traction stress.  As a result, the mobility of inclusion particles depend on the particular leaflet in which they are embedded.  In the large intermonolayer slip limit, the mobility is determined by a combination of these different solvent tractions from each of the leaflets.

When investigating the mobility of membrane inclusion particles, the finite spatial extent and curved geometry of the bilayer can result in important hydrodynamic effects not captured when treating the membrane as an infinite flat sheet.  We remark that the key consideration is how large the spatial extent or curvature is relative to the Saffman-Delbr\"uck length $L_{SD}$.  For a very large vesicle radius or small curvatures, we do expect of course to recover similar behaviours as in the case of an infinite flat sheet.  The interesting case is when the vesicle radius or membrane curvature yields a scale comparable or smaller than the Saffman-Delbr\"uck length $L_{SD}$.  

We show the self-mobility of an inclusion particle embedded in a membrane treated as an infinite flat sheet in Figure~\ref{fig:flatMembrane}.  These results were obtained by solving in Fourier space the hydrodynamic flow in response to an applied force density following closely the analytic approach presented in~\cite{Saffman1976, Oppenheimer2009} and our method for computing the mobility tensor discussed in Section~\ref{sec:mobilityOperators}.  We see significant differences compared to the mobility responses in spherical bilayers.

In the regime of a vesicle radius comparable to the the Saffman-Delbr\"uck length $L_{SD}$, the finite spatial extent of the membrane and topology can play an important role.  For spherical leaflets, it is required that mobility responses result in recirculation flows of the material within the finite leaflet.  As we have seen, this can yield non-trivial behaviours in the coupling and provide possibly useful flow features for estimating viscosity as discussed in Section~\ref{sec:vortices}.  

In contrast when treating the membrane as an infinite flat sheet, no vortex arises in the flows generated from single particle responses.  The infinite flat sheet also no longer results in trapped fluid within an interior domain. The bulk solvent fluid is treated as occupying an effectively infinite domain on both sides of the bilayer.  This results in more traction stress acting on the infinite flat sheet relative to the spherical shell which as a result reduces the self-mobility and strength of the coupled mobilities.  In particular, as the intramembrane viscosity increases the rotational mode of the spherical case is no longer available and the self-mobility decays to zero, see Figure~\ref{fig:flatMembrane}.  Our results show that significant differences can arise with treatment of the bilayer as an infinite flat sheet requiring treatment of the finite domain size and curved geometry of the bilayer when these length scales are comparable to the Saffman-Delbr\"uck length $L_{SD}$.  

\clearpage
\newpage 

\multicolinterrupt{
\begin{figure}[H]
\centering
\includegraphics[width=0.9\columnwidth]{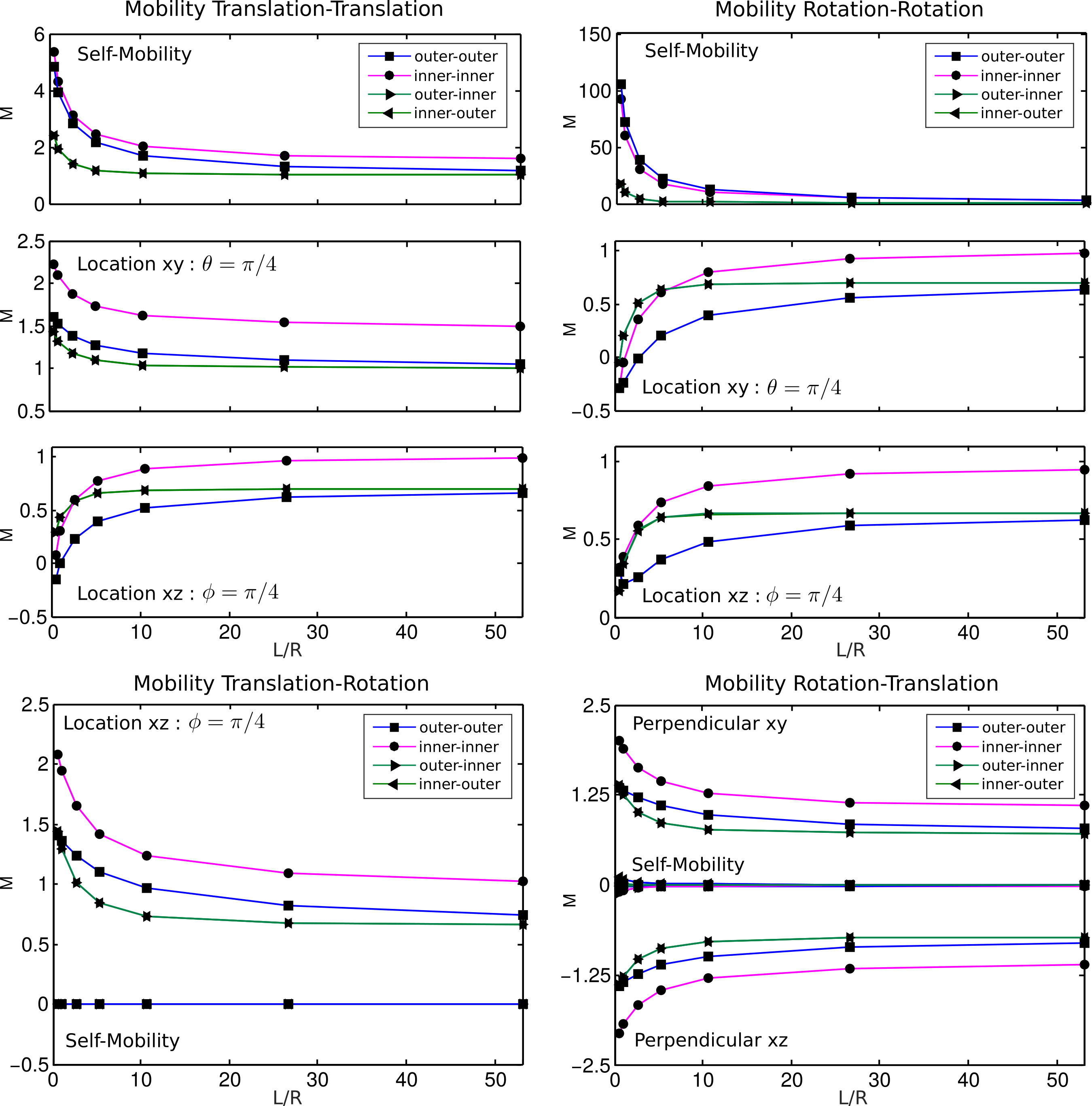}
\caption{Membrane Viscosity and Particle Mobility.  For a torque applied to a particle embedded within either the inner or outer leaflet, we show as the membrane viscosity is varied the translational and rotational responses of inclusion particles embedded within the inner or outer leaflet.  The intermonolayer slip is kept fixed at $\gammaMuRatio = \gamma/\gamma_0 = 4$.  This is discussed in more detail in Section ~\ref{sec:mobilityResults}. 
 \label{fig:viscAll}}
\end{figure}
}

\clearpage
\newpage 
.

\clearpage
\newpage

\multicolinterrupt{
\begin{figure}[H]
\centering
\includegraphics[width=0.99\columnwidth]{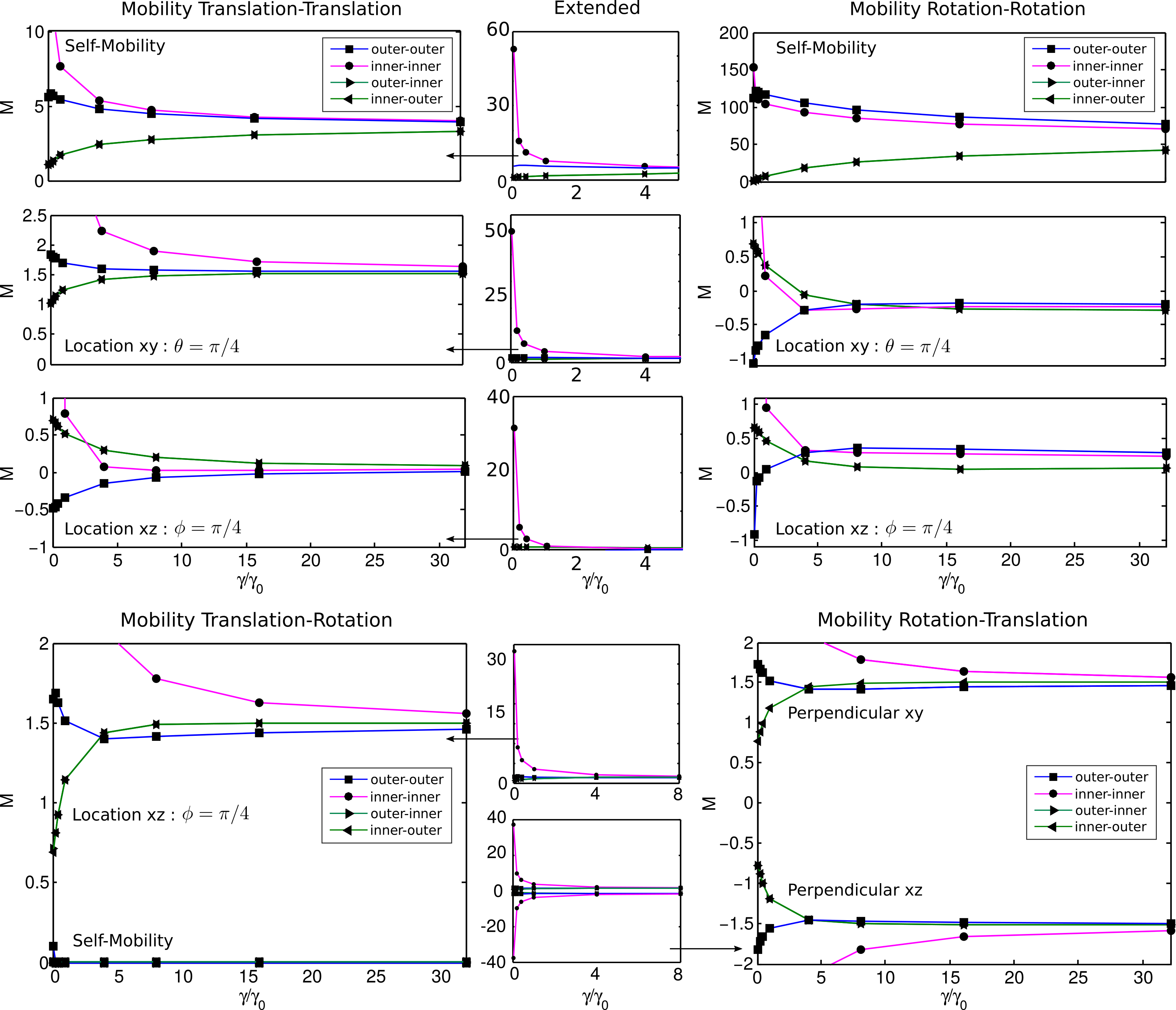}
\caption{Intermonolayer Slip and Particle Mobility.  For a torque applied to a particle embedded within either the inner or outer leaflet, we show as the intermonolayer slip is varied the translational and rotational responses of inclusion particles embedded within the inner or outer leaflet.  The membrane viscosity is kept fixed at $\LRratio = L/R = 0.13$.  This is discussed in more detail in Section ~\ref{sec:mobilityResults}. 
 \label{fig:slipAll}}
\end{figure}
}

\clearpage
\newpage 
\clearpage

\clearpage
\newpage

\subsection{Many-Particle Dynamics : Hydrodynamic Coupling and Diffusion}
\label{sec:manyParticleDynamics}

The collective dynamics of multiple particles within a spherical membrane can be modelled as 
\begin{eqnarray}
\label{equ_full_BD_model}
&& \frac{d\mb{X}}{dt} = \mb{M} \mb{F} +  k_B{T} \nabla\cdot \mb{M} + \mb{F}_{thm} \\
\nonumber
&& \langle \mb{F}_{thm}(s) \mb{F}_{thm}(t)^T\rangle = 2k_B{T}\mb{M}\delta(t - s).
\end{eqnarray}
The $\mb{X}$ denotes the collective particle configuration and $\mb{F}$ the collective forces acting on the particles.  The mobility $\mb{M}$ is obtained from the hydrodynamic-coupling methods introduced in Section~\ref{sec:mobilityTensor}.  The thermal fluctuations driving diffusion are accounted for by the drift $k_B{T} \nabla\cdot \mb{M}$ and the Gaussian random force $\mb{F}_{thm}(t)$ which is $\delta$-correlated in time with mean zero and covariance $\langle \mb{F}_{thm}(s) \mb{F}_{thm}(t)^T\rangle = 2k_B{T}\mb{M}\delta(t - s)$~\cite{AtzbergerTabak2015,AtzbergerSELM2011,Gardiner1985}.  The equations for the particles are to be interpreted in the sense of Ito Calculus~\cite{Oksendal2000a,Gardiner1985}.  The thermal drift term arises from the configuration-dependent correlations of the thermal fluctuations~\cite{AtzbergerTabak2015,AtzbergerSELM2011}.

\begin{figure}[H]
\centering
\includegraphics[width=0.8\columnwidth]{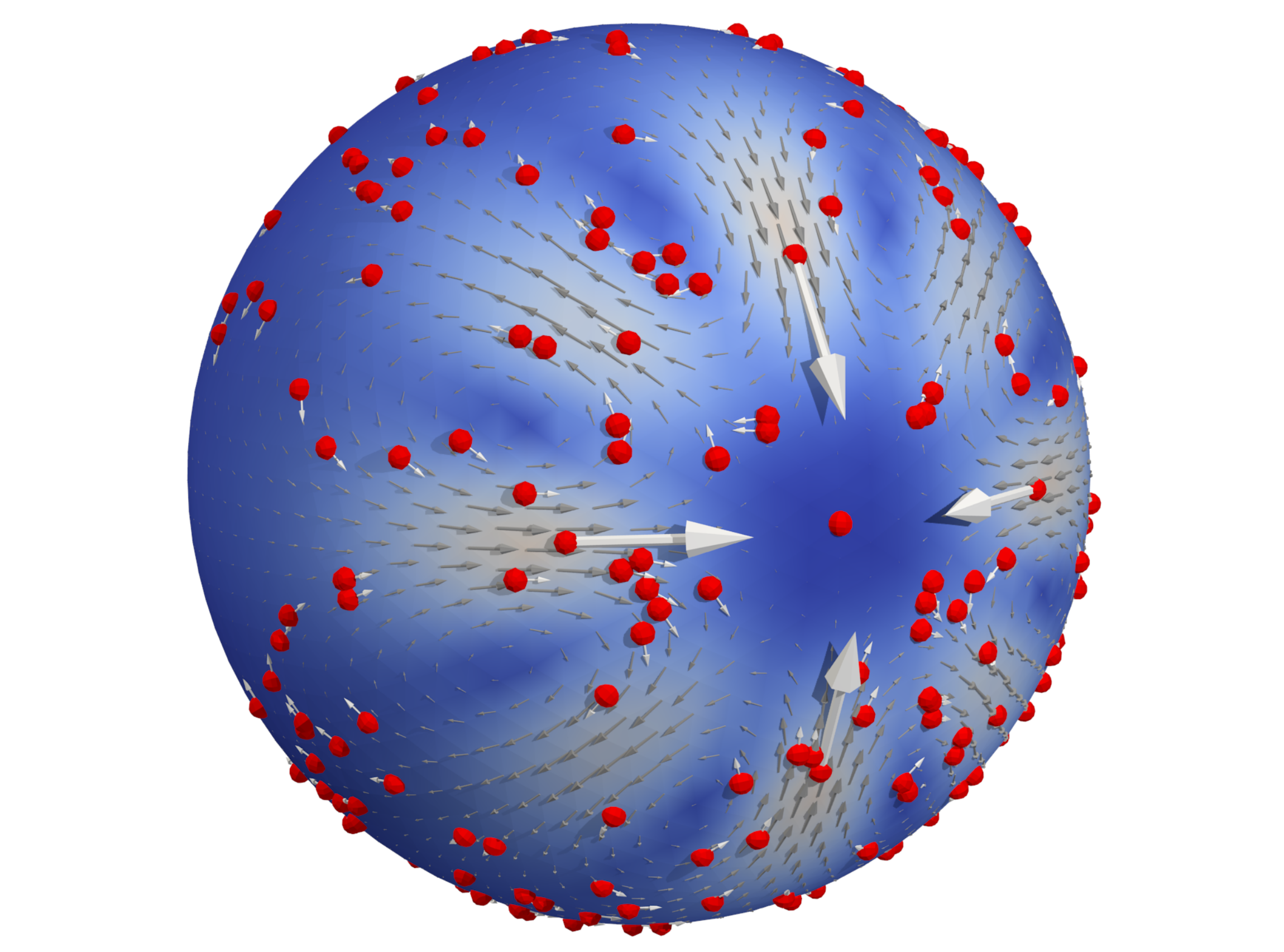}
\caption{Many-particle Dynamics within a Spherical Lipid Bilayer Membrane.  The inclusion particles are coupled through hydrodynamic flow both within the membrane bilayers and through the external solvent fluid. We show the hydrodynamic response in the case of a group of four inclusion particles attracted to a central particle.  We show the velocity of the other particles passively advected by the flow that either move in the opposite direction or are swept along depending on their relative location to the attracted particles.
 \label{fig:multiparticleConfig}}
\end{figure}

We focus here in this paper on how our approaches can be used for the collective hydrodynamics of particles within the membranes of spherical vesicles.  We defer to a future paper the full use of our introduced methods for the entire stochastic Brownian-hydrodynamic model in equation~\ref{equ_full_BD_model}.  As a demonstration of the introduced methods, we consider the specific case of $4$ particles that are actively attracted to a central particle located on the positive x-axis at the east pole.  We consider the hydrodynamic flow and particle dynamics within the outer-leaflet of the curved spherical bilayer.  In addition to the $4$ attracting particles, we also consider $195$ passive tracer particles that are advected by the flow, see Figure~\ref{fig:multiparticleConfig}. 

\begin{figure}[H]
\centering
\includegraphics[width=\columnwidth]{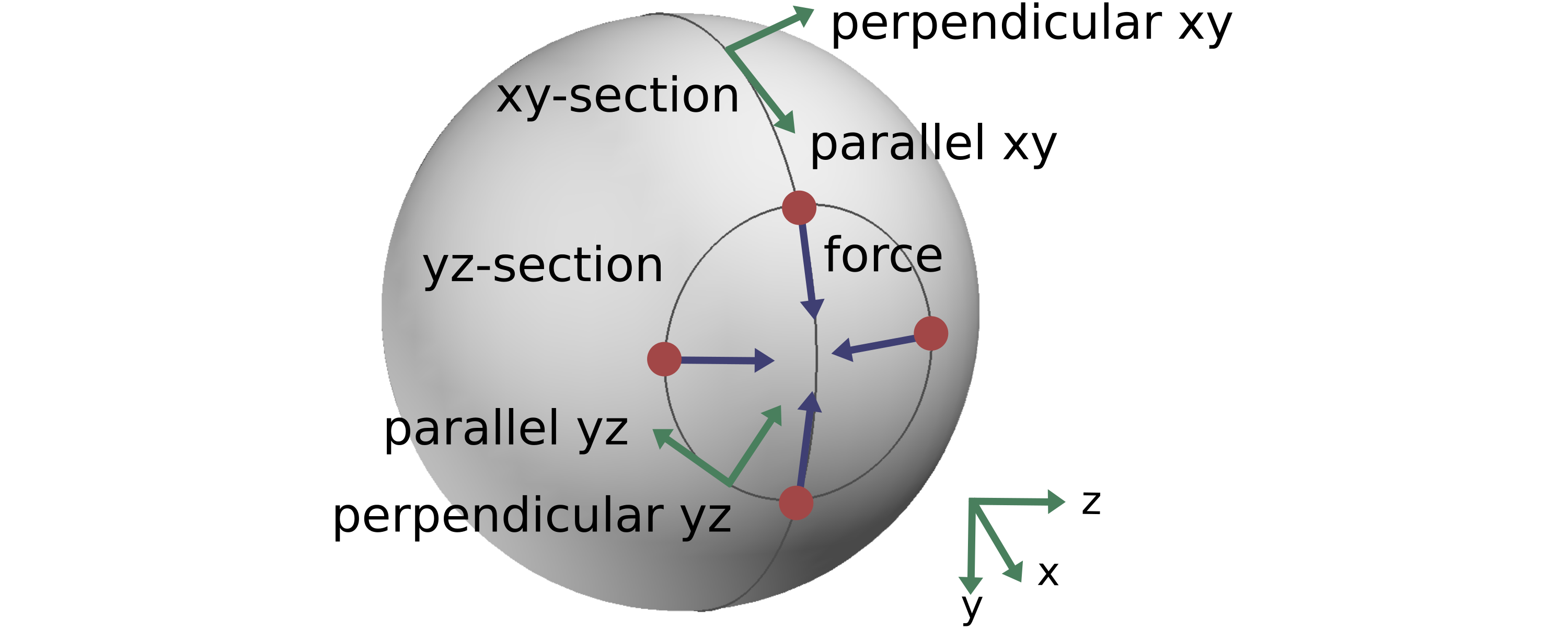}
 \caption{Cross Sections of the Sphere and Conventions.  We consider the hydrodynamic responses of the inclusion particles on two cross-sections of the sphere.  The first is the great circle of the sphere when intersected with the $xy$-plane.  The second is the circle of radius $r_0$ on the sphere surface parallel to the $xz$-plane.  For forces applied to the four attracting inclusion particles, we consider for the motions of the other inclusion particles as characterized by the mobility components parallel and perpendicular to the tangent of the respective cross-section curves.  We parametrise the $xz$-section using angle $\theta$ with $0$ corresponding to the location $(x,y,z) = (1,0,0)$  and the $xy$-section using angle $\theta$ with $0$ for location $(x,y,z) = (0,0,r_0)$.
 \label{fig:diagramCrossSectionsCircle}}
\end{figure}

We see that the hydrodynamic coupling can result in interesting dynamics with the passive particles either moving in the opposite direction of the attracting particles or swept along depending on their relative location.  This can be characterized by looking at the coupled mobility $M$ of the passively advected particles defined by $M = V/F_T$.  The $V$ is the passive particle velocity, $F_T$ the total force acting on the attracted particles.  We consider the responses in the circular section in the $yz$-plane of radius $r_0 = 0.5R$ centred about the x-axis near the east pole and in the circular section in the $xz$-plane of radius $r_0 = R$ about the center of the sphere, see  Figure~\ref{fig:multiparticleConfig} and Figure~\ref{fig:diagramCrossSectionsCircle}.  The parameters in these calculations are taken to be the same as in Table~\ref{table:defaultParams}.  

\begin{figure}[H]
\centering
\includegraphics[width=8cm]{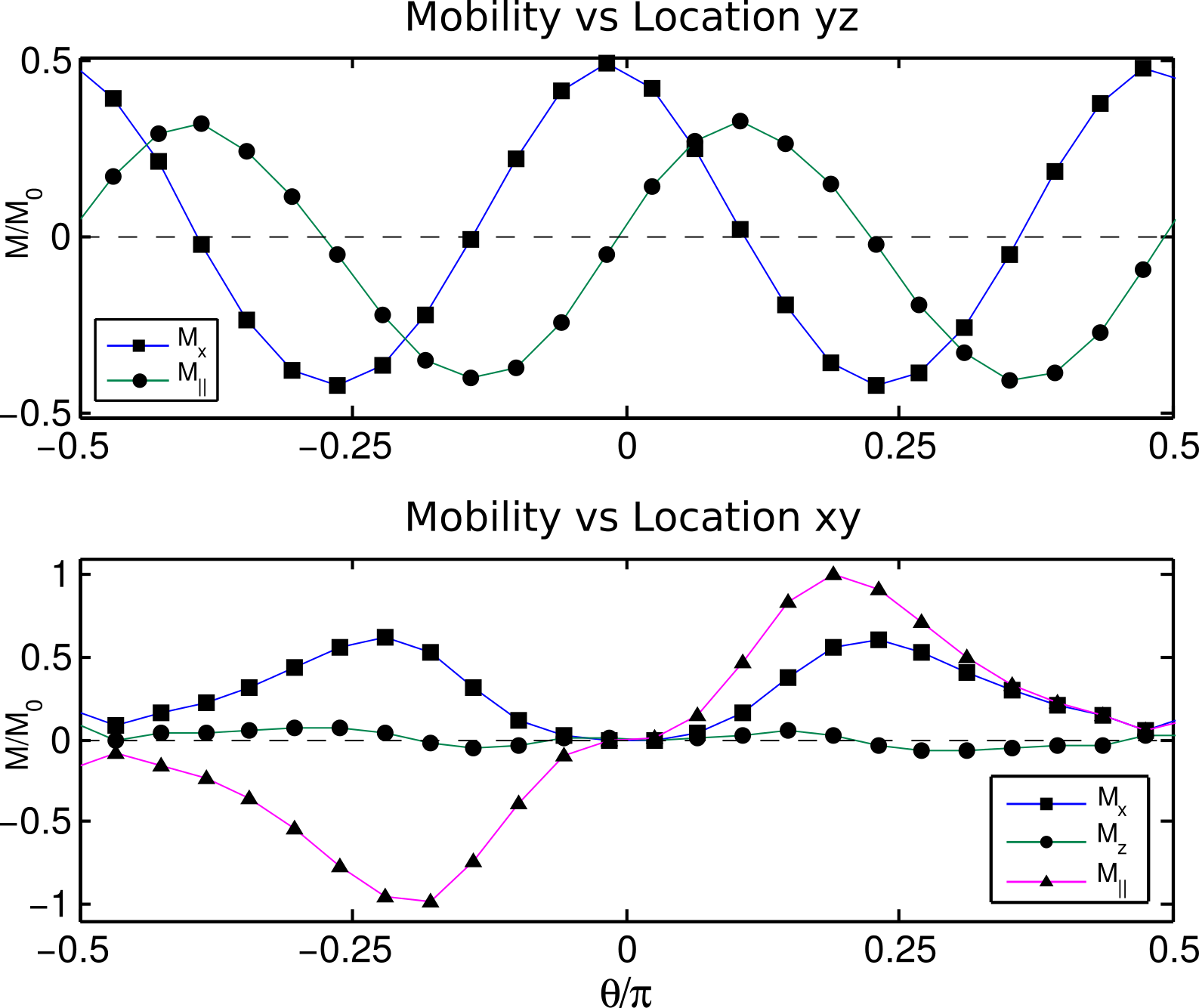}
\caption{Multi-particle Mobility.  We show the location dependent mobility of the passively advected particles in response  to the hydrodynamic coupling to the four attracting particles.  We show $M = V/F_T$ where $V$ is the particle velocity, $F_T$ the total force, 
$M_\parallel$ is the mobility tangent along the circular section.  The $yz$ indicates the circular section in the $yz$-plane of radius $r_0 = 0.5R$ about the east pole and $xz$ indicates the circular section in the $xz$-plane of radius $r_0 = R$ about the sphere center, see Figure~\ref{fig:diagramCrossSectionsCircle}.  In the response, depending on the position, the passive particles either move in the opposite direction or are swept along with the attracting particles.  The maximum response $M_0$ corresponds to the self-mobility of each of the attracting particles.
 \label{fig:multiparticleApp}}
\end{figure}

We see from the $yz$-responses $M_x$ that for locations half-way between the attracted particles, the passive particle move in the opposite direction to the attracted particles.  This change in direction is a consequence of the incompressibility of the fluid which results in an out-flow to compensate for the in-flow toward the east pole generated by the attracted particles, see  Figure~\ref{fig:multiparticleApp}.    
We also see this manifest in the $yz$-responses $M_{\parallel}$ which are out of phase with $M_x$ reflecting that the passive particles move laterally toward the out-flow half-way between the attracted particles.  The $xz$-responses correspond to passive particle motions when located on the same great circle in the $xz$-plane as two of the attracted particles.  We see in these responses that the passive particles always move toward the attracting particle at the east pole, see bottom panel of Figure~\ref{fig:multiparticleApp}.

These results indicate some of the rich dynamics that can arise from hydrodynamic coupling even for relatively simple configurations of particles and force laws.  The analytic approaches and computational methods we have introduced for the collective mobility tensor $M$ allow for incorporating such effects into simulations of many-particle dynamics within spherical lipid bilayers.  Many of the approaches we have introduced can also be extended for more general curved bilayers.

\section{Conclusions}
We have investigated the hydrodynamics of inclusion particles embedded within curved lipid bilayers.  We have performed an extensive study of the hydrodynamic flows and mobility responses within spherical bilayers.  We have studied both forces and torques applied to inclusion particles embedded within distinct inner and outer leaflets of the bilayer.  We have found significant differences relative to the case when the membrane is treated as a single infinite flat sheet.  We have found that the interplay between curvature, topology, and the two leaflet bilayer structure can yield interesting effects in the hydrodynamics.  We found for spherical bilayers that the difference between the infinite exterior solvent domain and the finite interior solvent domain result in different traction stresses acting on each of the leaflets.  We found this can have significant consequences for the mobility of inclusion particles especially when embedded within the inner leaflet.  We further found that the intermonolayer slip can play an interesting role with flow regimes where inclusion particles at the same location but in different leaflets can move in opposing directions in response to the lipid flows generated by other inclusion particles.  Our results show the rich individual and collective dynamics that can arise for inclusions within bilayers of spherical vesicles.  Many of our analytic approaches and computational methods also can be extended to study inclusions embedded within more general curved lipid bilayer membranes.

\section{Acknowledgements}
The authors P.J.A, J.K.S. acknowledge support from research grant
NSF CAREER DMS-0956210 and DOE ASCR CM4 DE-SC0009254.

\bibliographystyle{plain}
\bibliography{paperDatabase}{}

\begin{thebibliography}{10}

\bibitem{Abraham1988}
R.~Abraham, J.E. Marsden, and T.S. Rațiu.
\newblock {\em Manifolds, Tensor Analysis, and Applications}.
\newblock Number v. 75. Springer New York, 1988.

\bibitem{Acheson1990}
D.~J. Acheson.
\newblock {\em Elementary Fluid Dynamics}.
\newblock Oxford Applied Mathematics and Computing Science Series, 1990.

\bibitem{Alberts2007}
B.~Alberts, A.~Johnson, P.~Walter, J.~Lewis, M.~Raff, and K.~Roberts.
\newblock {\em Molecular Cell Biology of the Cell, 5th Ed.}
\newblock Garland Publishing Inc, New York, 2007.

\bibitem{ArroyoRelaxationDynamics2009}
Marino Arroyo and Antonio DeSimone.
\newblock Relaxation dynamics of fluid membranes.
\newblock {\em Phys. Rev. E}, 79(3):031915--, March 2009.

\bibitem{AtzbergerTabak2015}
P.~Atzberger and G.~Tabak.
\newblock Systematic stochastic reduction of inertial fluid-structure
  interactions subject to thermal fluctuations.
\newblock {\em SIAM J. Applied Mathematics (accepted)}, 2015.

\bibitem{Atzberger2006}
P.~J. Atzberger.
\newblock Velocity correlations of a thermally fluctuating brownian particle: A
  novel model of the hydrodynamic coupling.
\newblock {\em Physics Letters A}, 351(4-5):225--230--, 2006.

\bibitem{Atzberger2007c}
P.~J. Atzberger.
\newblock A note on the correspondence of an immersed boundary method
  incorporating thermal fluctuations with stokesian-brownian dynamics.
\newblock {\em Physica D-Nonlinear Phenomena}, 226(2):144--150--, 2007.

\bibitem{Atzberger2007a}
P.~J. Atzberger, P.~R. Kramer, and C.~S. Peskin.
\newblock A stochastic immersed boundary method for fluid-structure dynamics at
  microscopic length scales.
\newblock {\em Journal of Computational Physics}, 224(2):1255--1292--, 2007.

\bibitem{AtzbergerSELM2011}
Paul~J. Atzberger.
\newblock Stochastic eulerian lagrangian methods for fluid–structure
  interactions with thermal fluctuations.
\newblock {\em Journal of Computational Physics}, 230(8):2821--2837, April
  2011.

\bibitem{Ayton2006}
Gary~S. Ayton, J.~Liam McWhirter, and Gregory~A. Voth.
\newblock A second generation mesoscopic lipid bilayer model: Connections to
  field-theory descriptions of membranes and nonlocal hydrodynamics.
\newblock {\em The Journal of Chemical Physics}, 124(6):064906, 2006.

\bibitem{Tieleman2013}
W.F.~Drew Bennett and D.~Peter Tieleman.
\newblock Computer simulations of lipid membrane domains.
\newblock {\em Biochimica et Biophysica Acta (BBA) - Biomembranes},
  1828(8):1765--1776, August 2013.

\bibitem{Calvo1990}
N.~Calvo and O.H. Campanella.
\newblock A novel geometry for rheological characterization of viscoelastic
  materials.
\newblock {\em Rheologica Acta}, 29(4):323--331, July 1, 1990.

\bibitem{Camley2012}
B.~A. Camley and F.~L.~H. Brown.
\newblock Contributions to membrane-embedded-protein diffusion beyond
  hydrodynamic theories.
\newblock {\em Phys. Rev. E}, 85:061921, 2012.

\bibitem{Camley2013}
B.~A. Camley and F.~L.~H. Brown.
\newblock Diffusion of complex objects embedded in free and supported lipid
  bilayer membranes: role of shape anisotropy and leaflet structure.
\newblock {\em Soft Matter}, 9:4767--4779, 2013.

\bibitem{Capovilla2002}
R~Capovilla and J~Guven.
\newblock Stresses in lipid membranes.
\newblock {\em Journal of Physics A: Mathematical and General}, 35(30):6233--,
  2002.

\bibitem{Cooke2005}
Ira~R. Cooke, Kurt Kremer, and Markus Deserno.
\newblock Tunable generic model for fluid bilayer membranes.
\newblock {\em Phys. Rev. E}, 72(1):011506--, July 2005.

\bibitem{Tieleman1997}
S.~J.~Marrink D.~P.~Tieleman and H.~J.~C. Berendsen.
\newblock A computer perspective of membranes: molecular dynamics studies of
  lipid bilayer systems.
\newblock {\em Biochim. Biophys. Acta.}, 1331:235--270, 1997.

\bibitem{NelsonStatMechMem2004}
T~Piran David~Nelson, Steven~Weinberg.
\newblock {\em Statistical Mechanics of Membranes and Surfaces}.
\newblock World Scientific Publishing, 2004.

\bibitem{Deserno2009}
Markus Deserno.
\newblock Mesoscopic membrane physics: Concepts, simulations, and selected
  applications.
\newblock {\em Macromolecular Rapid Communications}, 30(9-10):752--771, 2009.

\bibitem{DesernoJanuary2015}
Markus Deserno.
\newblock Fluid lipid membranes: From differential geometry to curvature
  stresses.
\newblock {\em Membrane mechanochemistry: From the molecular to the cellular
  scale}, 185:11--45, January 2015.

\bibitem{Dimova1999}
R.~Dimova, C.~Dietrich, A.~Hadjiisky, K.~Danov, and B.~Pouligny.
\newblock Falling ball viscosimetry of giant vesicle membranes: Finite-size
  effects.
\newblock 12(4):589--598--, 1999.

\bibitem{Driscoll1994}
J.R. Driscoll and D.M. Healy.
\newblock Computing fourier transforms and convolutions on the 2-sphere.
\newblock {\em Advances in Applied Mathematics}, 15(2):202--250, June 1994.

\bibitem{Du2004}
Qiang Du, Chun Liu, and Xiaoqiang Wang.
\newblock A phase field approach in the numerical study of the elastic bending
  energy for vesicle membranes.
\newblock {\em Journal of Computational Physics}, 198(2):450--468, August 2004.

\bibitem{Farago2003}
Oded Farago.
\newblock ``water-free'' computer model for fluid bilayer membranes.
\newblock {\em J. Chem. Phys.}, 119(1):596--605, July 2003.

\bibitem{Klug2006}
Feng Feng and William~S. Klug.
\newblock Finite element modeling of lipid bilayer membranes.
\newblock {\em Journal of Computational Physics}, 220(1):394--408, December
  2006.

\bibitem{Gardiner1985}
C.~W. Gardiner.
\newblock {\em Handbook of stochastic methods}.
\newblock Series in Synergetics. Springer, 1985.

\bibitem{HappelBrenner1983}
J.~Happel and H.~Brenner.
\newblock {\em Low Reynolds Number Hydrodynamics: With Special Applications to
  Particulate Media}.
\newblock Springer Netherlands, 1983.

\bibitem{LevineDinsmoreHydroEffectTopology2008}
M.~L. Henle, R.~McGorty, A.~B. Schofield, A.~D. Dinsmore, and A.~J. Levine.
\newblock The effect of curvature and topology on membrane hydrodynamics.
\newblock {\em EPL (Europhysics Letters)}, 84(4):48001--, 2008.

\bibitem{LevineHenleHydroCurvedMembranes2010}
Mark~L. Henle and Alex~J. Levine.
\newblock Hydrodynamics in curved membranes: The effect of geometry on
  particulate mobility.
\newblock {\em Phys. Rev. E}, 81(1):011905--, January 2010.

\bibitem{HonerkampSmith2013}
Aurelia~R. Honerkamp-Smith, Francis~G. Woodhouse, Vasily Kantsler, and
  Raymond~E. Goldstein.
\newblock Membrane viscosity determined from shear-driven flow in giant
  vesicles.
\newblock {\em Phys. Rev. Lett.}, 111(3):038103--, July 2013.

\bibitem{Lowengrub2007}
J.-J.~Xu J.~Lowengrub and A.~Voigt.
\newblock Surface phase separation and flow in a simple model of multicomponent
  drops and vesicles.
\newblock {\em Fluid Dyn. Mater. Proc. v. 3}, 1-19, 2007.

\bibitem{Jarvis2004}
Tyler Jarvis and James Tanton.
\newblock The hairy ball theorem via sperner's lemma.
\newblock {\em The American Mathematical Monthly}, 111(7):599--603, August
  2004.

\bibitem{Muller2012}
Osman Kahraman, Norbert Stoop, and Martin~Michael Müller.
\newblock Fluid membrane vesicles in confinement.
\newblock {\em New Journal of Physics}, 14(9):095021--, 2012.

\bibitem{Lamb1895}
H.~Lamb.
\newblock {\em Hydrodynamics}.
\newblock University Press, 1895.

\bibitem{Lebedev1999}
V.~I. Lebedev and D.~N. Laikov.
\newblock A quadrature formula for the sphere of the 131st algebraic order of
  accuracy.
\newblock {\em Dokl. Math.}, 59:477–481, 1999.

\bibitem{LevineViscoelastic2002}
A.~J. Levine and F.~C. MacKintosh.
\newblock Dynamics of viscoelastic membranes.
\newblock {\em Phys. Rev. E}, 66:061606, 2002.

\bibitem{LevineMobilityExtendedBodies2004}
Alex~J. Levine, T.~B. Liverpool, and F.~C. MacKintosh.
\newblock Mobility of extended bodies in viscous films and membranes.
\newblock {\em Phys. Rev. E}, 69(2):021503--, February 2004.

\bibitem{Marsden1994}
J.E. Marsden and T.J.R. Hughes.
\newblock {\em Mathematical Foundations of Elasticity}.
\newblock Dover, 1994.

\bibitem{Naji2007c}
A.~Naji, A.~J. Levine, and P.~A. Pincus.
\newblock Corrections to the saffman-delbr{\"u}ck mobility for membrane bound
  proteins.
\newblock {\em Biophys. J.}, 93:L49--L51, 2007.

\bibitem{Noguchi2004}
Hiroshi Noguchi and Gerhard Gompper.
\newblock Fluid vesicles with viscous membranes in shear flow.
\newblock {\em Phys. Rev. Lett.}, 93(25):258102--, December 2004.

\bibitem{Chou2008}
Sarah~A. Nowak and Tom Chou.
\newblock Membrane lipid segregation in endocytosis.
\newblock {\em Phys. Rev. E}, 78(2):021908--, August 2008.

\bibitem{Oksendal2000a}
B.~Oksendal.
\newblock {\em Stochastic Differential Equations: An Introduction}.
\newblock Springer, 2000.

\bibitem{Oppenheimer2009}
Naomi Oppenheimer and Haim Diamant.
\newblock Correlated diffusion of membrane proteins and their effect on
  membrane viscosity.
\newblock {\em Biophysical Journal}, 96(8):3041--3049, April 2009.

\bibitem{Groves2007}
Raghuveer Parthasarathy and Jay~T. Groves.
\newblock Curvature and spatial organization in biological membranes.
\newblock {\em Soft Matter}, 3(1):24--33, 2007.

\bibitem{Peskin2002}
Charles~S. Peskin.
\newblock The immersed boundary method.
\newblock {\em Acta Numerica}, 11:1--39, July 2002.

\bibitem{Powers2002}
Thomas~R. Powers, Greg Huber, and Raymond~E. Goldstein.
\newblock Fluid-membrane tethers: Minimal surfaces and elastic boundary layers.
\newblock {\em Phys. Rev. E}, 65(4):041901--, March 2002.

\bibitem{Pressley2001}
A.~Pressley.
\newblock {\em Elementary Differential Geometry}.
\newblock Springer, 2001.

\bibitem{AtzbergerBassereau2014}
François Quemeneur, Jon~K. Sigurdsson, Marianne Renner, Paul~J. Atzberger,
  Patricia Bassereau, and David Lacoste.
\newblock Shape matters in protein mobility within membranes.
\newblock {\em Proceedings of the National Academy of Sciences},
  111(14):5083--5087, 2014.

\bibitem{OsterSteigmann2013}
Padmini Rangamani, Ashutosh Agrawal, KranthiK. Mandadapu, George Oster, and
  DavidJ. Steigmann.
\newblock Interaction between surface shape and intra-surface viscous flow on
  lipid membranes.
\newblock 12(4):833--845--, 2013.

\bibitem{Reynwar2007}
Benedict~J. Reynwar, Gregoria Illya, Vagelis~A. Harmandaris, Martin~M. Muller,
  Kurt Kremer, and Markus Deserno.
\newblock Aggregation and vesiculation of membrane proteins by
  curvature-mediated interactions.
\newblock {\em Nature}, 447(7143):461--464, May 2007.

\bibitem{Saffman1976}
P.~G. Saffman.
\newblock Brownian motion in thin sheets of viscous fluid.
\newblock {\em J. Fluid Mech.}, 73:593--602, 1976.

\bibitem{Saffman1975}
P.~G. Saffman and M.~Delbr\"{u}ck.
\newblock Brownian motion in biological membranes.
\newblock {\em Proc. Natl. Acad. Sci. USA}, 72:3111--3113, 1975.

\bibitem{Vlahovska2011}
Jonathan~T. Schwalbe, Petia~M. Vlahovska, and Michael~J. Miksis.
\newblock Vesicle electrohydrodynamics.
\newblock {\em Phys. Rev. E}, 83:046309, Apr 2011.

\bibitem{Seifert1997}
U.~Seifert.
\newblock Configurations of fluid membranes and vesicles.
\newblock {\em Advances in Physics}, 46(1):13--137--, 1997.

\bibitem{Seifert1993}
U.~Seifert and S.~A. Langer.
\newblock Viscous modes of fluid bilayer membranes.
\newblock {\em Europhys. Lett.}, 23:71--76, 1993.

\bibitem{AtzbergerSigurdsson2012}
Jon~K. Sigurdsson, Frank~L.H. Brown, and Paul~J. Atzberger.
\newblock Hybrid continuum-particle method for fluctuating lipid bilayer
  membranes with diffusing protein inclusions.
\newblock {\em Journal of Computational Physics}, 252(0):65--85, November 2013.

\bibitem{Voeltz2007}
Gia~K. Voeltz and William~A. Prinz.
\newblock Sheets, ribbons and tubules [mdash] how organelles get their shape.
\newblock {\em Nat Rev Mol Cell Biol}, 8(3):258--264, March 2007.

\bibitem{Woodhouse2012}
Francis~G. Woodhouse and Raymond~E. Goldstein.
\newblock Shear-driven circulation patterns in lipid membrane vesicles.
\newblock {\em Journal of Fluid Mechanics}, 705:165--175, 2012.

\end{thebibliography}

\clearpage
\newpage

\appendix
\noindent
\section*{Appendix} 
\section{Spherical Harmonic Methods : Lebedev Quadratures, SPH Transform, and Polar Singularities}
\label{sec:coord_charts}
\begin{figure}[H]
\centering
\includegraphics[width=8cm]{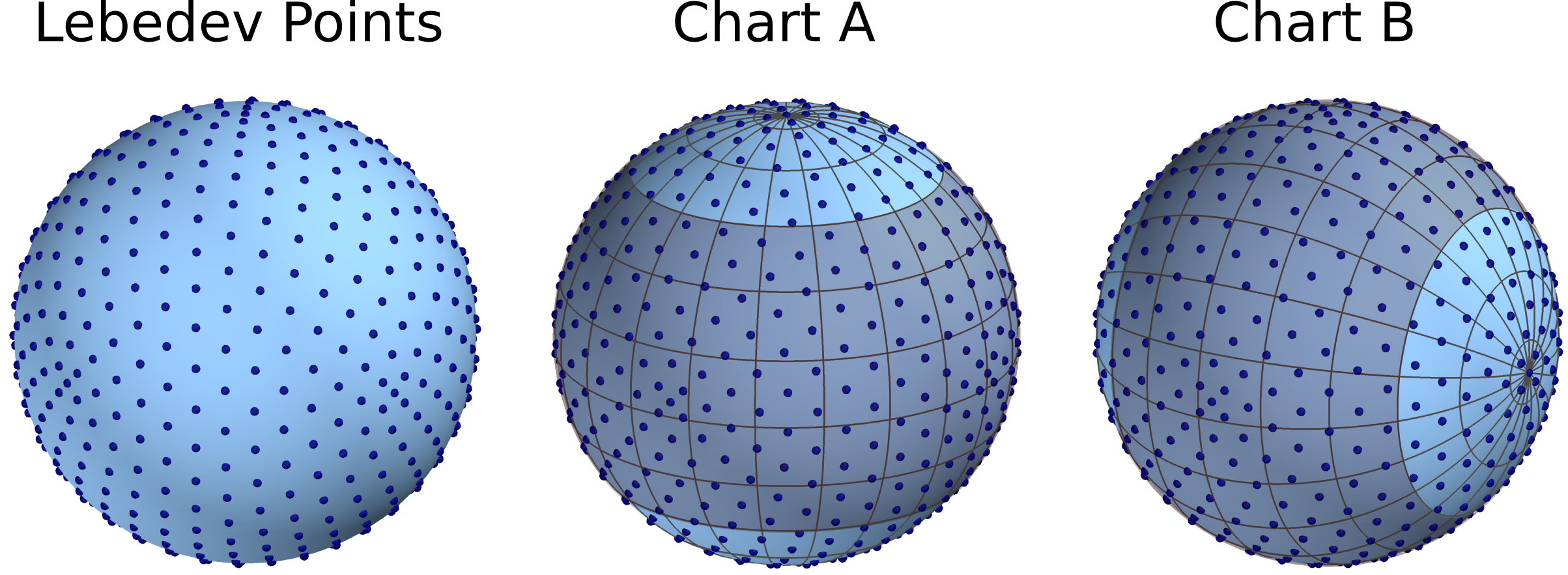}
\caption{Locations of the 590 Lebedev quadrature points and the two charts $A$ and $B$ used to deal with singularities at the poles.
\label{fig:lebedevSchematic}}
\end{figure}

We make a few brief remarks on our methods for numerical computations and the issues that arise when performing calculations on spherical surfaces. We have developed our methods using high-order Lebedev quadratures which integrate exactly spherical harmonics up to large degree~\cite{Lebedev1999}. We evaluate inner-products using 
\begin{eqnarray}
\nonumber
\langle f,g \rangle_\mathcal{S} = \int_{\mathcal{S}} f(\mb{x})g(\mb{x}) dA = \sum_{m}  w_m f(\mb{x}_m) g(\mb{x}_m).
\end{eqnarray}
The $w_m$ are the weights and $\mb{x}_m$ the nodes.
We use this to compute spherical harmonic coefficients by the inner-product $\hat{f}_s = \langle f,\Phi_s \rangle$, where $\Phi_s$ is the spherical harmonic with index $s = (m,\ell)$.  While computationally more expensive than Fast Spherical Harmonic Transforms (FSHT)~\cite{Driscoll1994}, a distinct of advantage of our Lebedev-based methods over the lattitude-longitude sampling of FSHT is the more uniform and symmetric sampling of Lebedev nodes which have octahedral symmetry~\cite{Lebedev1999}, see Figure~\ref{fig:lebedevSchematic}.

On a spherical surface, coordinate singularities arise, such as exhibited by spherical coordinates at the north-south pole.  We handle this issue by using two alternative coordinate charts $A$ or $B$ depending on the location.  Chart $A$ corresponds to the spherical coordinates with singularities at the north and south poles.  Chart $B$ corresponds to the spherical coordinates with singularities at the west and east poles.  A notable feature of the Lebedev sampling is that its symmetry allows us to make use of the same quadrature nodes in the two coordinate charts, see Figure~\ref{fig:lebedevSchematic}.

To perform numerical calculations for operations on field values at the Lebedev points, such as divergence, gradient, and curl, we make use whenever possible of the intrinsic geometric meaning of such operations (as opposed to the coordinate-centric formulas).  When coordinate-centric formulas are used, we express the operations in one of the two different coordinate charts $A$ or $B$. The particular chart is chosen to yield significant distance to the coordinate-system singularities.  This allows for robust numerical calculations at all points on the sphere surface.  This further highlights one of the advantages of our less coordinate-centric exterior calculus approach to the hydrodynamics.

We obtain the membrane velocity field in our calculations by
\begin{eqnarray}
\mb{v} & = & \left(\mb{v}^{\flat}\right)^{\sharp} = \sum_s a_s \left(-\star \mb{d} \Phi_s\right)^{\sharp} \\
\nonumber
& = & \sum_s a_s
\left[ \frac{\epsilon_{i\ell}}{\sqrt{|g|}} \frac{\partial \Phi}{\partial x^{\ell}} \right]\partial_{\mb{x}^{i}}.
\end{eqnarray}
The $|g| = \mbox{det}(\mb{g})$ is the determinant of the metric tensor and $\epsilon_{i\ell}$ is the Levi-Civita tensor (slight abuse of notation).  The $x^\ell$ denotes the coordinates.  For spherical coordinates in chart $A$, $x^1 = \theta^A, x^2 = \phi^A$ and in chart $B$, $x^1 = \theta^B, x^2 = \phi^B$.  
To obtain the velocity, we express in each of the charts the coordinate derivatives of the spherical harmonic modes $\partial{\Phi}/{\partial x^{\ell}}$ and the basis vectors $\partial_{\mb{x}^i}$ in terms of the embedding space $\mathbb{R}^3$.  We then choose at each given location the expression for the chart that has a significant distance to the coordinate-system singularities.  In this manner, we compute robustly the velocity field over the entire surface.

\end{multicols}

\end{document}